\documentclass[aps,pra,twocolumn,superscriptaddress]{revtex4-2}
\usepackage{amsmath,amsfonts,amssymb}
\usepackage{appendix}
\usepackage{graphicx,float,calc}
\usepackage{color,bm}
\usepackage{ulem}
\usepackage{braket}\hyphenpenalty=5000
\usepackage[colorlinks,urlcolor=blue,citecolor=blue,linkcolor=blue]{hyperref}

\begin{document}
	
\title{Asymmetric and chiral dynamics of two-component anyons with synthetic gauge flux}
	
\author{Rui-Jie Chen}
\affiliation{Key Laboratory of Atomic and Subatomic Structure and Quantum Control (Ministry of Education), Guangdong Basic Research Center of Excellence for Structure and Fundamental Interactions of Matter, South China Normal University, Guangzhou 510006, China}
\affiliation{Guangdong Provincial Key Laboratory of Quantum Engineering and Quantum Materials,School of Physics, South China Normal University, Guangzhou 510006, China}

\author{Ying-Xin Huang}
\affiliation{Key Laboratory of Atomic and Subatomic Structure and Quantum Control (Ministry of Education), Guangdong Basic Research Center of Excellence for Structure and Fundamental Interactions of Matter, South China Normal University, Guangzhou 510006, China}
\affiliation{Guangdong Provincial Key Laboratory of Quantum Engineering and Quantum Materials,School of Physics, South China Normal University, Guangzhou 510006, China}

\author{Guo-Qing Zhang}
\affiliation{Research Center for Quantum Physics, Huzhou University, Huzhou 313000, People's Republic of China}

\author{Dan-Wei Zhang}
\email{danweizhang@m.scnu.edu.cn}
\affiliation{Key Laboratory of Atomic and Subatomic Structure and Quantum Control (Ministry of Education), Guangdong Basic Research Center of Excellence for Structure and Fundamental Interactions of Matter, South China Normal University, Guangzhou 510006, China}
\affiliation{Guangdong Provincial Key Laboratory of Quantum Engineering and Quantum Materials,School of Physics, South China Normal University, Guangzhou 510006, China}	
\affiliation{Quantum Science Center of Guangdong-Hong Kong-Macao Greater Bay Area (Guangdong), Shenzhen 518045, China}

\date{\today}	
	
\begin{abstract}
In this work, we investigate the non-equilibrium dynamics in a one-dimensional two-component anyon-Hubbard model, which can be mapped to an extended Bose-Hubbard ladder with density-dependent hopping phase and synthetic gauge flux. Through numerical simulations of few-particle dynamics and the analytical symmetry analysis without particle-number constraints, we reveal the asymmetric and imbalance transports with broken inversion symmetry and find two dynamical symmetries in the expansion dynamics unique to the two-component anyon-Hubbard chain. For any initial states that preserves both the time-reversal and inversion symmetries, the expansion is always dynamically symmetric under spatial inversion and component flip, when the sign of anyonic statistics phase or both the signs of gauge flux and interaction are changed. For vanishing Hubbard interaction, we show the dynamical suppression induced by both the statistics phase and the gauge flux. For finite Hubbard interaction, we further uncover tunable chiral-antichiral dynamics for two-component anyons under the gauge flux. We demonstrate that both chiral and antichiral dynamics can be exhibited and tuned by the statistics phase and the gauge flux. The dynamical phase regimes with respect to the chiral-antichiral dynamics for different system parameters are obtained. These findings highlight the novel dynamical phenomena arising from the interplay of anyonic exchange statistics, synthetic gauge fields, and Hubbard interactions in multi-component anyons.
\end{abstract}	

\maketitle

\section{\label{intro}Introduction}
In three dimensions, fundamental particles can be classified as either bosons or fermions according to the 0 or $\pi$ phase acquired upon exchanging two identical particles. There exists other quantum statistics for quasiparticles in two dimensions, where the exchange phase can interpolate between the bosonic and fermionic limits and such quasiparticles are called anyons~\cite{Leinaas1977,Wilczek1982,Laughlin1983,Wilczek2024}. In two-dimensional fractional quantum Hall systems, there are two types of anyonic quasiparticles: Abelian anyons that acquire a phase factor upon exchange, and non-Abelian 
appear where quasiparticle exchange yields unitary matrices, leading to a state transformation on a degenerate ground-state manifold. Anyons have become an important concept in various areas of modern physics, ranging from the fractional quantum Hall effect~\cite{Laughlin1983,Halperin1984,Kim2005} and quantum spin liquids~\cite{Coldea2001,Kitaev2006,Yao2007,Bauer2014,Semeghini2021} to topological quantum computation~\cite{Kitaev2003,Sarma2005,Nayak2008,Lindner2013,Google2023}. The concept of fractional statistics for Abelian anyons has been generalized from two dimensions to arbitrary dimensions by Haldane~\cite{Haldane1991}, which is reformulated as a generalization of the Pauli exclusion principle. In particular, Abelian anyons in one dimension have recently attracted increasing attentions~\cite{Ha1994,Murthy1994,Yue1995,Amico1998,Kundu1999,Batchelor2006,Girardeau2006,Hao2008,Hao2009,Hao2012,Greiter2009,Zinner2015,Alexey2018,Zhang2020,Linhu2024}. Some intriguing properties of one-dimensional anyons have been revealed, such as asymmetric momentum distribution~\cite{Greiter2009,Hao2008,Hao2009,Hao2012}, statistically induced quantum phase transition~\cite{Keilmann2011,Greschner2015,Arcila2016,Zuo2018,Agarwala2019,Bonkhoff2025}, spatially asymmetric particle transport~\cite{Alexey2018}, and anyonic symmetry protected topological phases~\cite{Lange2017}. 
The momentum distributions and off-diagonal correlations of the many-body ground states of one-dimensional anyonic Lieb-Liniger gases in the impenetrable limit has been systematically and analytically studied in Refs.~\cite{Santachiara2008,Calabrese2009,Scopa2020}.
The non-equilibrium dynamics, spectral functions, and dynamical bosonization of these strongly interacting anyons have also been investigated \cite{OIP2020,OIP2022,OIP2025}.

Several schemes have been proposed for the creation and manipulation of anyonic statistics in ultracold atomic systems, such as two-dimensional topological optical superlattice \cite{Paredes2008} and one-dimensional optical lattices with the Raman-assisted tunneling~\cite{Keilmann2011,Greschner2015} or Floquet driving ~\cite{Greschner2014,Cardarelli2016,Christoph2016,Yuan2017,Greschner2018,Frederik2019,Lienhard2020,Greiner2024}. 
By engineering four-body ring-exchange interactions within an ultracold atom system, a minimal toric-code Hamiltonian \cite{Kitaev2003} has been implemented to observe the fractional statistics of Abelian anyons~\cite{Dai2017}. Recently, one-dimensional anyons have been realized in a two-body setting using ultracold bosons in a tilted optical lattice subjected to proper Floquet modulations~\cite{Greiner2024}. This system provides access to a continuum of statistical phases and enables the observation of the asymmetric transport, in contrast to the symmetric dynamics of bosons and fermions. The one-dimensional anyons have also been realized in a strongly interacting quantum gas with an impurity via the spin-charge separation \cite{Dhar2025,BWang2025}, where the asymmetric momentum distribution and dynamical fermionization were observed. However, most of  current theoretical and experimental studies on one-dimensional anyonic systems are predominantly confined to single-component anyons, leaving the properties of multi-component anyons with internal degrees of freedom largely unexplored.

On the other hand, growing effort has been made to engineer synthetic gauge fields with ultracold atoms ~\cite{Dalibard2011,Goldman2014,Goldman2016,DW2018,Cooper2019,Lin2009,Aidelsburger2013,Miyake2013,Yan2022,JWang2025a,Zhu2006,Zhu2011,DWZhang2020}. For instance, the simplest system for studying the dynamical phenomena under synthetic gauge fields is two-leg ladders with effective magnetic fields~\cite{Dariol2014,Atala2014,Tai2017,Bryce2017,dutt2020single,Yan2024,Yan2025EngineeringTC,YLChen2020,GQZhang2021}. The chiral currents and vortex phases have been observed in bosonic ladders under synthetic fluxes, both in the absence and presence of interatomic interactions~\cite{Atala2014,Tai2017,Piraud2015,Zheng2017,Li2020,Giri2023,Yan2024}. Moreover, the antichiral dynamics have been revealed in several non-Hermitian two-leg ladder systems~\cite{Wu2022,Chen2024,Ye2025}, where particles on both legs propagate along the same direction. The two-leg ladder lattice can be engineered in real space~\cite{Atala2014,Tai2017}, or by employing the two internal states of atoms as a synthetic dimension~\cite{Bryce2017Direct,dutt2020single,Suotang2022,Yan2024}. Yet, most of the studies related to two-leg flux ladders focus on bosons or fermions, and the effect of anyonic statistics on the quantum dynamics in these ladders remains unclear.

In this work, we propose a one-dimensional two-component anyon-Hubbard model, which can be mapped to an extended Bose-Hubbard ladder with density-dependent hopping phase and synthetic gauge flux. We investigate the novel non-equilibrium dynamics in this two-component anyon-Hubbard chain under the interplay of anyonic statistics, synthetic gauge flux, and interactions. Based on numerical simulations of few-particle dynamics and the analytical symmetry analysis without particle-number constraints, we reveal the spatially asymmetric and component-imbalance transports and two hidden dynamical symmetries in the expansion dynamics of two-component anyons. For vanishing Hubbard interaction, we show that the time-reversal symmetry is broken under the fractional statistics phases and non-zero gauge fluxes, while inversion symmetry is always preserved. For finite Hubbard interaction, the time-reversal symmetry is restored only when the gauge flux is absent, whereas the inversion symmetry is maintained only when both the statistics phase and gauge flux are 0 or $\pi$. For any initial states that preserves both the time-reversal and inversion symmetries, we uncover that the expansion of two-component anyons is dynamically symmetric under spatial inversion and component flip, when the sign of statistics phase or the signs of both gauge flux and Hubbard interaction are changed. For vanishing Hubbard interaction, we also show that both the statistics phase and gauge flux suppress the expansion. Furthermore, for finite Hubbard interaction, we demonstrate that both chiral and antichiral dynamics can be exhibited and tunable by the statistics phase and gauge flux. The dynamical phase regimes with respect to the chiral-antichiral dynamics for different system parameters are obtained.

The rest of this paper is organized as follows. In Sec.~\ref{model}, we introduce the two-component anyon-Hubbard model with synthetic gauge flux and establish its mapping to a bosonic counterpart. Sec.~\ref{sec_dysy} is devoted to revealing the asymmetric expansion dynamics and dynamical symmetries. In Sec.~\ref{diff}, we demonstrate the dynamical suppression and chiral-antichiral dynamics under the statistics phase and gauge flux. A brief discussion on the four-particle dynamics and a short conclusion are given in Sec.~\ref{conclusion}. The derivation details on dynamical symmetries are presented in the Appendix ~\ref{App}.

\section{\label{model}Model}

\begin{figure}[t!]
	\centering
	\includegraphics[width=0.48\textwidth]{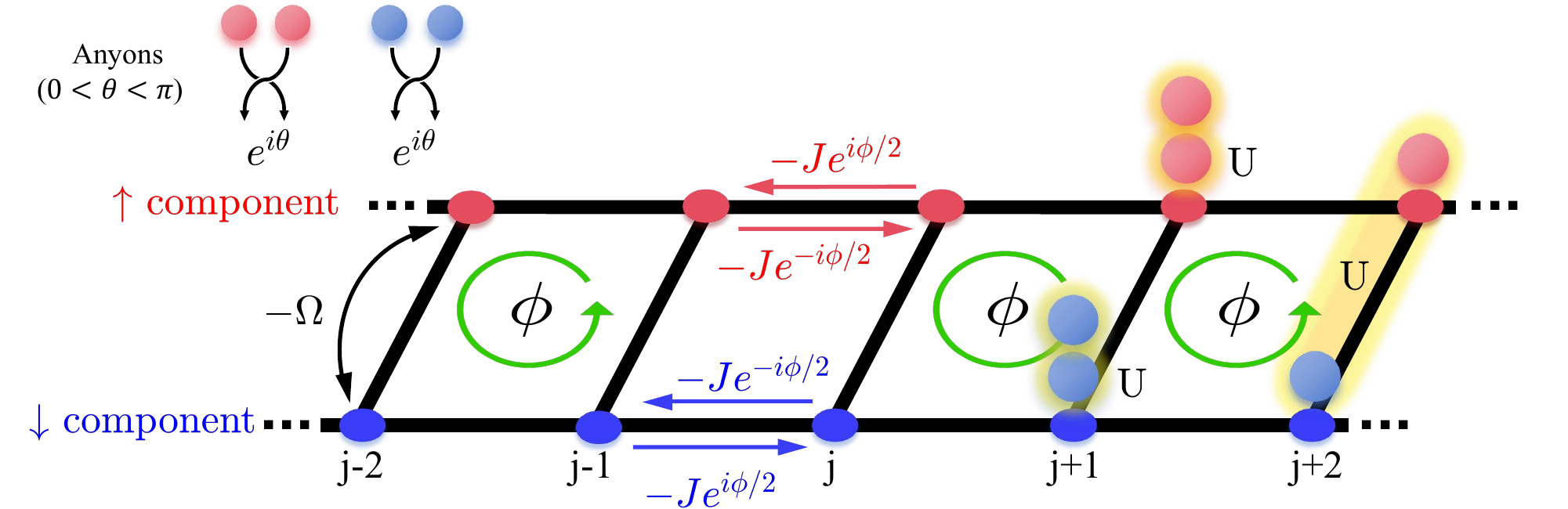}
	\caption{Schematic of the lattice of interacting two-component anyons under an artificial magnetic flux denoted by $\phi$. Here, $J$ denotes the hopping amplitude, $U$ is the Hubbard interaction strength, and $\Omega$ represents the coupling strength between two components. One-dimensional anyons with two components have an exchange phase $\theta$ that interpolates between 0 and $\pi$.}
	\label{fig1}
\end{figure}

We consider a one-dimensional two-component anyon-Hubbard model, as illustrated in Fig. \ref{fig1}. The model is described by the following Hamiltonian 
\begin{equation}\label{HamA}
	\begin{split}					\hat{H}_A=&-J\sum_{j,\sigma}^{L-1}(\hat{a}^\dagger_{j,\sigma}\hat{a}_{j+1,\sigma}e^{i\phi_{\sigma}/2}+\text{H.c.})\\&-\Omega\sum_{j}^{L}(\hat{a}^\dagger_{j,\uparrow}\hat{a}_{j,\downarrow}+\text{H.c.})+\frac{U}{2}\sum_{j}^{L}\hat{n}_j(\hat{n}_j-1).
	\end{split}
\end{equation}
Here, $\hat{a}_{j,\sigma}^{\dagger}$ ($\hat{a}_{j,\sigma}$) creates (annihilates) an anyon at site $j$ with components $\sigma=\{\uparrow,\downarrow\}$, $\hat{n}_j=\hat{n}_{j,\uparrow}+\hat{n}_{j,\downarrow}$ with $\hat{n}_{j,\sigma}=\hat{a}_{j,\sigma}^{\dagger}\hat{a}_{j,\sigma}$ is the particle number operator, and $L$ denotes the lattice size. $J$ denotes the intra-component hopping amplitude, $\Omega$ represents the coupling strength between two components, and $U$ denotes the on-site Hubbard interaction strength. Note that even for the vanishing Hubbard interaction with $U=0$, the anyons are already intrinsically interacting. Hereafter, we set $J=1$ as the energy unit and $\hbar/J$ as the time unit.

This system can be regarded as an anyonic two-leg ladder with synthetic gauge field~\cite{Giri2023}, where two components correspond to two legs, as shown in Fig.~\ref{fig1}. Each plaquette is threaded by a synthetic magnetic flux $\phi$, arising from the hopping phase factor $e^{i\phi_{\sigma}/2}$ with $\phi_{\uparrow}=\phi$ and $\phi_{\downarrow}=-\phi$. 
When $\phi = 0$, our model reduces to two coupled one-dimensional anyon-Hubbard chains. In a single-component anyon-Hubbard chain, the asymmetric transport induced by anyonic statistics has been revealed in Ref. \cite{Alexey2018}. Here we extend the asymmetric transport and related dynamical symmetries to two-component anyons, and further explore the interplay among anyonic statistics, synthetic gauge flux, and Hubbard interactions  on novel non-equilibrium dynamics in this ladder system.

The anyonic creation and annihilation operators obey the generalized commutation relations ~\cite{Wilczek1982,Keilmann2011}
\begin{equation}\label{comm_1}
	\begin{aligned}		&[\hat{a}_{j,\sigma},\hat{a}^{\dagger}_{l,\sigma'}]_{\theta}\equiv\hat{a}_{j,\sigma}\hat{a}^{\dagger}_{l,\sigma'}-e^{-i\theta \mathrm{sgn}(j-l)}\hat{a}^{\dagger}_{l,\sigma'}\hat{a}_{j,\sigma}=\delta_{j,l}\delta_{\sigma,\sigma'},\\		&[\hat{a}_{j,\sigma},\hat{a}_{l,\sigma'}]_{\theta}\equiv\hat{a}_{j,\sigma}\hat{a}_{l,\sigma'}-e^{i\theta \mathrm{sgn}(j-l)}\hat{a}_{l,\sigma'}\hat{a}_{j,\sigma}=0,
	\end{aligned}	
\end{equation}
where $\theta$ is the exchange statistics phase and $\mathrm{sgn}(j-l)$ takes values $-1,0,1$ for $j<l$, $j=l$, $j>l$, respectively. When $\theta=0$, particles behave as the same as bosons. When $\theta=\pi$, the two-component anyons are pseudofermionic \cite{Hao2008}: they behave as fermions on different sites, yet multi particles can occupy the same site, similar to bosons. Exchanging anyons with the same component between different sites give rise to an additional phase factor in the many-body wave function. We can map the one-dimensional anyons to bosons by using the fractional Jordan-Wigner transformation~\cite{Keilmann2011,Alexey2018} $\hat{a}_{j,\sigma}=\hat{b}_{j,\sigma}e^{i\theta\sum_{l<j}\hat{n}_{l}}$
and $\hat{a}_{j,\sigma}^{\dagger}=e^{-i\theta\sum_{l<j}\hat{n}_{l}}\hat{b}_{j,\sigma}^{\dagger}$, with $\hat{b}_{j,\sigma}^{\dagger}$ ($\hat{b}_{j,\sigma}$) being the boson creation (annihilation) operator. We obtain the corresponding two-component extended Bose-Hubbard model
\begin{equation}\label{HamB}
	\begin{split}			\hat{H}_B=&-J\sum_{j,\sigma}(\hat{b}^\dagger_{j,\sigma}e^{i(\theta\hat{n}_{j}+\phi_{\sigma}/2)}\hat{b}_{j+1,\sigma}+\text{H.c.})\\&-\Omega\sum_{j}(\hat{b}^\dagger_{j,\uparrow}\hat{b}_{j,\downarrow}+\text{H.c.})+\frac{U}{2}\sum_{j}^{L}\hat{n}_j(\hat{n}_j-1),
	\end{split}
\end{equation}
where $e^{i\theta\hat{n}_{j}}$ represents an occupation-dependent Peierls phase with $\hat{n}_j=\hat{n}_{j,\uparrow}+\hat{n}_{j,\downarrow}$, and $\hat{n}_{j,\sigma}=\hat{b}_{j,\sigma}^{\dagger}\hat{b}_{j,\sigma}=\hat{a}_{j,\sigma}^{\dagger}\hat{a}_{j,\sigma}$.
Thus, one can simulate the two-component anyon-Hubbard model in Eq. (\ref{HamA}) by using this extended Bose-Hubbard model in Eq. (\ref{HamB}). The required density-dependent phase $e^{i\theta\hat{n}_{j}}$ and the component-dependent hopping phase factor $e^{i\phi_{\sigma}/2}$ have been individually realized for bosonic atoms in one-dimensional optical lattices under Floquet and Raman engineering hoppings, respectively~\cite{Greiner2024,Atala2014,Tai2017}.

We study the expansion dynamics of two-component anyons initially localized at the central region of the lattice. To reveal the dynamical symmetry and asymmetric dynamics, we require the initial states preserving both the inversion and time-reversal symmetries (see Appendix.~\ref{App} for details). For simplicity and convenience in our simulations, we consider the initial state of anyons as a product state in Fock space, $|\psi_0\rangle_A=\prod_j\hat{a}_{j,\sigma}^{\dagger}|\mathrm{vac}\rangle$,  occupied symmetrically around the lattice center. Here $|\mathrm{vac}\rangle$ denotes the vacuum state. Under the fractional Jordan-Wigner transformation, the initial state picks up an irrelevant global phase factor $e^{i\beta}$, i.e., $|\psi_0\rangle_A=e^{i\beta}\prod_j\hat{b}_{j,\sigma}^{\dagger}|\mathrm{vac}\rangle=|\psi_0\rangle_B$. Consequently, the anyonic and bosonic particle densitis at the time $t$ are related via
\begin{equation}
  \langle\hat{n}_{j,\sigma}^A(t)\rangle={}_B\langle\psi_0|e^{i\hat{H}_Bt}\hat{n}_{j,\sigma}e^{-i\hat{H}_Bt}|\psi_0\rangle_B=\langle\hat{n}_{j,\sigma}^B(t)\rangle.
\end{equation}
This indicates that the particle densities for anyons and bosons are equivalent in the time evolution. Thus, we can perform numerical simulations and symmetry analysis based on the extended Bose-Hubbard Hamiltonian in Eq. (\ref{HamB}), and omit the indices $A$ and $B$ for simplicity.

The main numerical results in this work are obtained via the exact diagonalization for simulating the two-particle dynamics from an initial product state $|\psi_0\rangle_A=1/\sqrt{2}(\hat{a}_{j_0,\uparrow}^{\dagger}\hat{a}_{j_0+1,\uparrow}^{\dagger}+\hat{a}_{j_0,\downarrow}^{\dagger}\hat{a}_{j_0+1,\downarrow}^{\dagger})|\mathrm{vac}\rangle$, where $j_0$ and $(j_0+1)$ are two central sites of the chain of length $L=26$. However, the asymmetric and imbalance transports and the dynamical symmetries revealed from the numerical results are not specific to this two-particle initial product state. Instead, they are general for any initial states that preserve the inversion and time-reversal symmetries, as analytically demonstrated from our symmetry analysis without particle-number constraints. To further confirm the dynamical (a)symmetries and other novel dynamics in this two-component anyon-Hubbard chain, we also simulate the four-particle dynamics by using the time-dependent variational principle algorithm \cite{Fishman2022}. In our numerical simulations, we use the open boundary condition and focus on the dynamics within the timescale that the particles do not reach the boundaries, such that we can identify the particle propagation from the time-evolved density distribution and ignore the boundary effect on the dynamics.

\section{\label{sec_dysy}Asymmetric transport and dynamical symmetries}

In this section, we investigate the expansion dynamics and its dynamical symmetries of two-component anyons. It has been shown that the exchange statistics of single-component interacting anyons gives rise to asymmetric transport~\cite{Alexey2018,Greiner2024}. On the other hand, the synthetic gauge flux and interaction leads to chirality in the two-boson dynamics~\cite{Tai2017,Giri2023}. The two dynamical phenomena imply the breakdown of inversion and time-reversal symmetries, respectively. For two-component anyons in our system, we further explore the dynamical symmetries and the asymmetric and imbalance transports under the interplay of the exchange statistics phase, the synthetic gauge flux, and the Hubbard interaction. 

\begin{figure}[t!]
	\centering
	\includegraphics[width=0.48\textwidth]{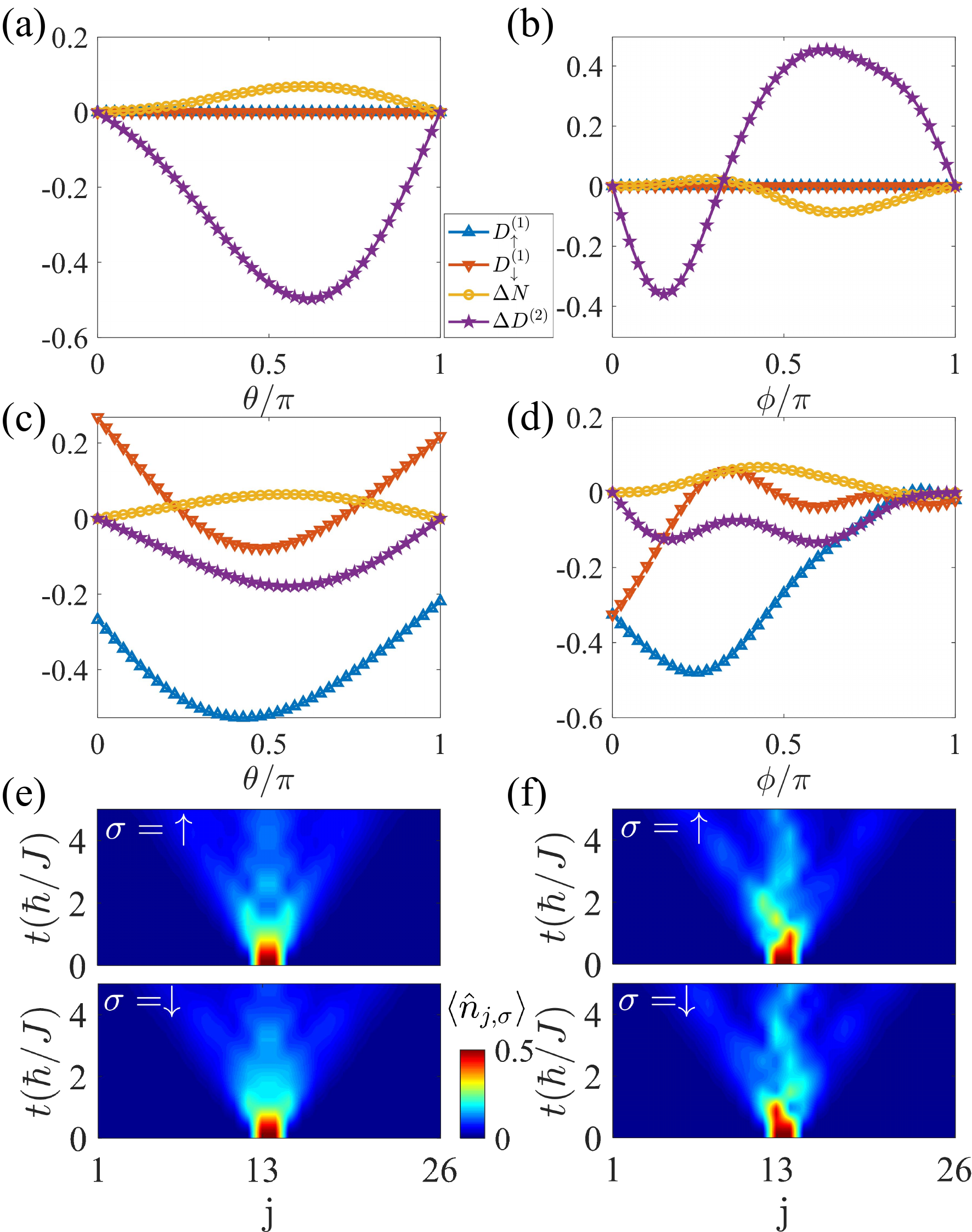}
	\caption{(Color online) (a-d) $D^{(1)}_{\uparrow}$, $D^{(1)}_{\downarrow}$, $\Delta N$ and $\Delta D^{(2)}$ as a function of $\theta$ or $\phi$ at the time $t=2$ (in units of $\hbar/J$). The non-interacting and interacting cases with $U/J=0$ and are $U/J=4$ are shown in (a,b) and (c,d), respectively. We set $\phi=\pi/4$ in (a,c) and $\theta=\pi/4$ in (b,d). Time evolution of density distributions for (e) $U/J=0$ and (f) $U/J=4$ with fixed $\theta=\pi/4$ and $\phi=\pi/2$. Other parameters in (a-f) are $\Omega/J=1$ and $L=26$.}
	\label{fig2}
\end{figure}

\subsection{Numerical results} 

We consider several physical quantities to numerically study the expansion dynamics and dynamical symmetries. The first quantity is the center-of-mass of the $\sigma$-component:
\begin{equation}	D^{(1)}_{\sigma}(t)=\sum_{j}^{L}(j-j_0)\langle\hat{n}_{j,\sigma}(t)\rangle.
\end{equation} 
A finite value of $D^{(1)}_{\sigma}$ after the time evolution indicates spatially asymmetric transport with broken inversion symmetry. Another quantity is the second momentum
\begin{equation}	D^{(2)}_{\sigma}(t)=\sum_{j}^{L}(j-j_0)^2\langle\hat{n}_{j,\sigma}(t)\rangle,
\end{equation}
which quantifies the spreading breadth. The third one is the component density imbalance 
\begin{equation}
\Delta N(t)=\sum_{j}^{L}\langle\hat{n}_{j,\uparrow}(t)\rangle-\langle\hat{n}_{j,\downarrow}(t)\rangle.
\end{equation}
Exact diagonalization results on the expansion dynamics from the two-particle initial state for various parameters in a system of size $L=26$ are shown in Fig. \ref{fig2}.

We first consider the vanishing Hubbard interaction case with $U=0$. Figs.~\ref{fig2}(a) and \ref{fig2}(b) show the results of $D^{(1)}_{\uparrow}$, $D^{(1)}_{\downarrow}$, $\Delta N$ and $\Delta D^{(2)}=D^{(2)}_{\uparrow}-D^{(2)}_{\downarrow}$ as a function of the statistics phase $\theta$ and the flux $\phi$ at the time $t=2$ (in units of $\hbar/J$), respectively. We find $D^{(1)}_{\uparrow}=D^{(1)}_{\downarrow}$, $\Delta N=0$ and $\Delta D^{(2)}=0$ only when $\theta$ or $\phi$ equals to 0 or $\pi$. This implies that in the case of $U=0$, the expansion dynamics preserve time-reversal symmetry only for bosons ($\theta=0$) or pseudofermions ($\theta=\pi$), or when the synthetic gauge flux is absent ($\phi=0$ or $\pi$). Otherwise, for a fractional statistics phase with $0<\theta<\pi$ or non-zero flux with $0<\phi<\pi$, the time-reversal symmetry is broken in the expansion dynamics. On the other hand, $D^{(1)}_{\uparrow}=D^{(1)}_{\downarrow}=0$ for any $\theta$ and $\phi$ indicates that the inversion symmetry always preserves in the expansion for $U=0$.

For finite Hubbard interaction with $U\neq0$, as shown in Figs.~\ref{fig2}(c) and \ref{fig2}(d), $D^{(1)}_{\uparrow}=D^{(1)}_{\downarrow}$, $\Delta N=0$ and $\Delta D^{(2)}=0$ at the time $t=2$ only when $\phi=0$ or $\pi$. This implies that a non-zero flux breaks the time-reversal symmetry in the interacting expansion dynamics. Moreover, $D^{(1)}_{\uparrow}\neq0$ or $D^{(1)}_{\downarrow}\neq 0$ whenever $\theta$ and $\phi$ take values other than 0 or $\pi$. Thus, the inversion symmetry is broken in the presence of fractional statistics phase and flux for $U\neq0$. Typical time evolutions of density distributions for $\theta=\pi/4$ and $\phi=\pi/2$ for $U=0$ and $U\neq0$ are shown in Figs.~\ref{fig2}(e) and \ref{fig2}(f), respectively. The asymmetric expansion exhibits for both two components of anyons for $U\neq0$, which is absent when $U=0$, similar to the asymmetric dynamics of the single-component anyons in Refs. \cite{Alexey2018,Greiner2024}. In addition, the two-component anyons exhibit component-imbalance transport for both vanishing and finite Hubbard interactions (see Figs.~\ref{fig2}(e) and \ref{fig2}(f)).

\subsection{Symmetry analysis and dynamical symmetries}

\begin{figure}[t!]
	\centering
	\includegraphics[width=0.48\textwidth]{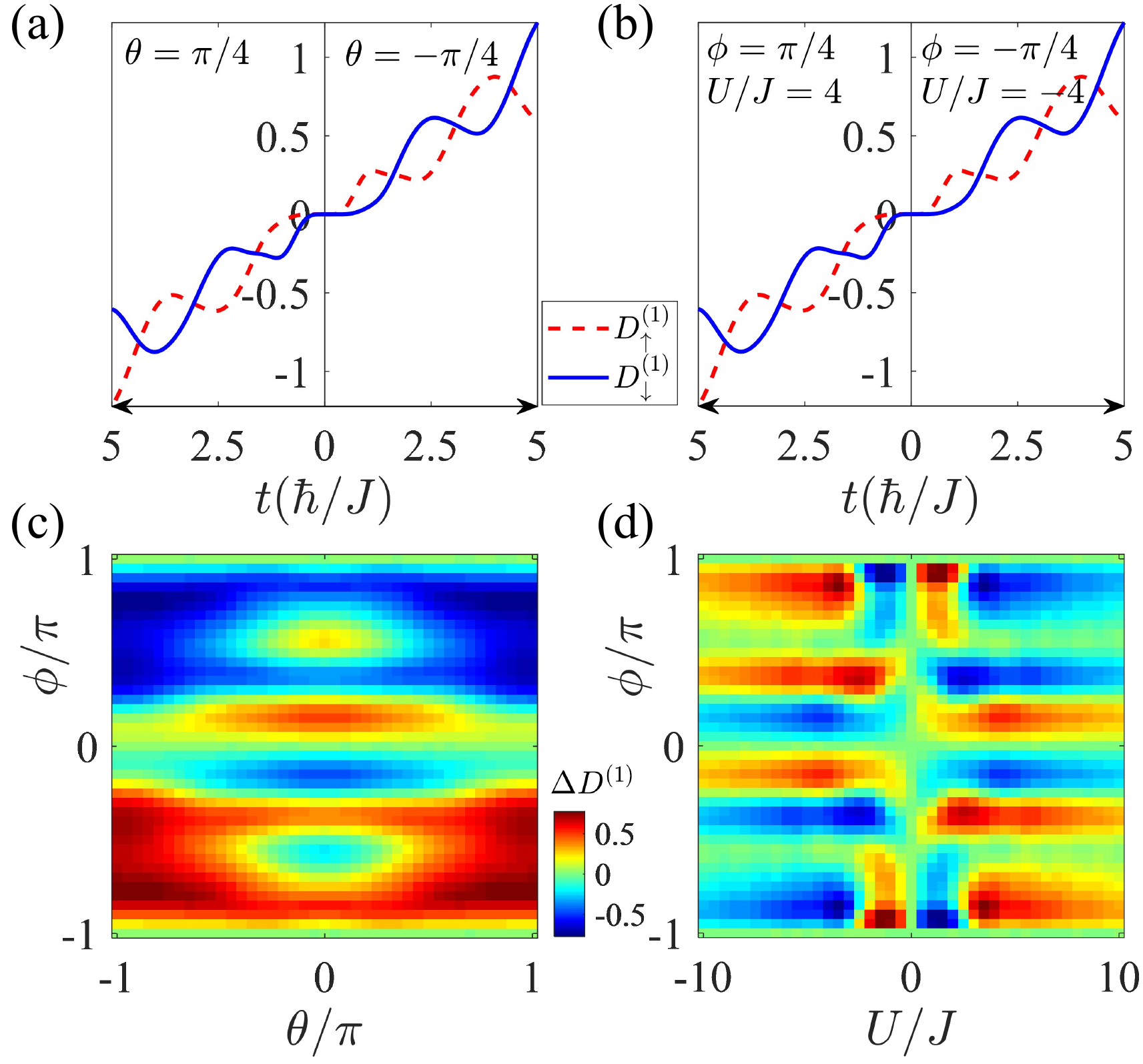}
	\caption{(Color online) Time evolution of $D^{(1)}_{\uparrow}$ and $D^{(1)}_{\downarrow}$ for (a) $\theta=\pm\pi/4$ and (b) $\phi=\pm\pi/4$ and $U/J=\pm4$. $\Delta D^{(1)}$ in (c) the $\theta$-$\phi$ plane and (d) the $U$-$\phi$ plane at the time $t=4$ (in units of $\hbar/J$). Other parameters are $\Omega=1$ and $L=26$ in (a-e), $\phi=\pi/4$ and $U/J=4$ in (a), $\theta=\pi/4$ in (b), $U/J=4$ in (c), and $\theta=\pi/4$ in (d).}
	\label{fig3}
\end{figure}

We proceed to perform symmetry analysis to reveal asymmetric dynamics and the related dynamical symmetries \cite{Schneider2012,Yu2017,Alexey2018}. 
As the occupation-dependent Peierls phase depends only on the occupation of the left site, the model Hamiltonian in Eq. (\ref{HamB}) generally breaks the inversion symmetry $\mathcal{I}$: $\mathcal{I}\hat{b}_{j,\sigma}\mathcal{I}^{\dagger}=\hat{b}_{j',\sigma}$, where $j$ and $j'$ are under reflection about the center of the chain. Under non-vanishing synthetic gauge flux, it also breaks the time-reversal symmetry $\mathcal{T}$: $\mathcal{T}\hat{b}_{j,\uparrow}\mathcal{T}^{-1}=\hat{b}_{j,\downarrow}$ and $\mathcal{T}i\mathcal{T}^{-1}=-i$. We consider a generic initial state $|\psi_0\rangle$ that preserves both the time-reversal and inversion symmetries, such as the two-particle initial state in our simulations. To characterize the symmetries in the expansion dynamics, we consider two target states that are related by the time-reversal symmetry $|\psi_1\rangle=\mathcal{T}|\psi_2\rangle$, and another state $|\psi_3\rangle$ that is related to $|\psi_1\rangle$ by the inversion symmetry $|\psi_1\rangle=\mathcal{I}|\psi_3\rangle$. Whether the expansion dynamics is time-reversal or inversion symmetric can be determined from the overlap between the final state and these target states $|\psi_d\rangle$ with $d=1,2,3$: $S_d=|\langle\psi_{d}|e^{-i\hat{H}_Bt}|\psi_0\rangle|$. If $S_1=S_2$ for any evolution time $t$, the time-reversal symmetry is preserved in the dynamics, otherwise it is broken. Similarly, the inversion symmetry in the dynamics is preserved if $S_1=S_3$ and broken otherwise. Based on the analysis in the Appendix~\ref{App}, we find that in the case of $U=0$, $S_1 = S_2$ for the time-reversal symmetry only when $\theta$ or $\phi$ equals 0 or $\pi$, whereas $S_1=S_3$ for the inversion symmetry holds for all values of $\theta$ and $\phi$. For the case of $U\neq0$, we find that $S_1=S_2$ for the time-reversal symmetry only when $\phi=0$ or $\pi$, while $S_1=S_3$ for the inversion symmetry only when both $\theta$ and $\phi$ are 0 or $\pi$ simultaneously. These analytical results confirm the numerical observations of the expansion dynamics shown in Fig.~\ref{fig2}.

Furthermore, we reveal the dynamical symmetries with respect to the parameters $\{U,\theta,\phi\}$. We define a symmetry operator $\mathcal{K}=\mathcal{R}\mathcal{I}\mathcal{T}$ with the density-dependent gauge transformation $\mathcal{R}=e^ {i\theta[\sum_{j,\sigma}^{L}\hat{n}_{j,\sigma}(\hat{n}_{j,\sigma}-1)/2+\sum_{j}^{L}\hat{n}_{j,\uparrow}\hat{n}_{j,\downarrow}]}$. 
Under the operator $\mathcal{R}$, the density dependent phase of the nearest neighbor hopping terms is modified, such as $\mathcal{R}\hat{b}_{j,\sigma}^{\dagger}e^{i(\theta\hat{n}_j+\phi_{\sigma}/2)}\hat{b}_{j+1,\sigma}\mathcal{R}^{-1}=\hat{b}_{j,\sigma}^{\dagger}e^{i(\theta\hat{n}_{j+1}+\phi_{\sigma}/2)}\hat{b}_{j+1,\sigma}$.
By combining the inverse operator $\mathcal{I}$ and the time-reversal operator $\mathcal{T}$, we find that the extended Bose-Hubbard Hamiltonian given by Eq. (\ref{HamB}) has the following property under the operator $\mathcal{K}$:
\begin{equation}\label{KHK}
	\mathcal{K}\hat{H}_B(\phi)\mathcal{K}^{\dagger}=\hat{H}_B(-\phi).
\end{equation}
Base on the symmetry analysis in Appendix.~\ref{App}, we find that the Hamiltonian $\hat{H}_B$ with opposite signs of system parameters $\{U,\theta,\phi\}$ is related by the transformations given by Eqs.(\ref{P},\ref{T},\ref{R}). The first one is
\begin{equation}\label{P}
    \mathcal{P}\hat{H}_B(U)\mathcal{P}^{\dagger}=-\hat{H}_B(-U),
\end{equation}
where $\mathcal{P}=e^{i\pi[\sum_{r}^{L/2}(\hat{n}_{2r+1,\uparrow}+\hat{n}_{2r+1,\downarrow})+\sum_{j}^{L}\hat{n}_{j,\uparrow}]}$ is the parity operator.
Under $\mathcal{P}$, the signs of the nearest neighbor hopping and the inter-component terms are reversed, while the Hubbard interaction term remains unchanged. For instance, one has $\mathcal{P}J\hat{b}^{\dagger}_{j,\sigma}e^{i(\theta\hat{n}_{j}+\phi_{\sigma}/2)}\hat{b}_{j+1,\sigma}\mathcal{P}^{\dagger}=-J\hat{b}^{\dagger}_{j,\sigma}e^{i(\theta\hat{n}_{j}+\phi_{\sigma}/2)}\hat{b}_{j+1,\sigma}$, $\mathcal{P}\Omega\hat{b}^{\dagger}_{j,\uparrow}\hat{b}_{j,\downarrow}\mathcal{P}^{\dagger}=-\Omega\hat{b}^{\dagger}_{j,\uparrow}\hat{b}_{j,\downarrow}$, and $\mathcal{P}\frac{U}{2}\hat{n}_j(\hat{n_j}-1)\mathcal{P}^{\dagger}=-\frac{U}{2}\hat{n}_j(\hat{n_j}-1)$. The second relation is given by 
\begin{equation}\label{T}
	\mathcal{T}\hat{H}_B(\theta)\mathcal{T}^{-1}=\hat{H}_B(-\theta),
\end{equation}
where the time-reversal operator $\mathcal{T}$ leads to the change of the hopping terms as $\mathcal{T}\hat{b}^{\dagger}_{j,\uparrow}e^{i\theta\hat{n}_{j}+i\phi/2}\hat{b}_{j+1,\uparrow}\mathcal{T}^{-1}=\hat{b}^{\dagger}_{j,\downarrow}e^{-i\theta\hat{n}_{j}-i\phi/2}\hat{b}_{j+1,\downarrow}$. 
When $\theta=0$ or $\pi$, the Hamiltonian $\hat{H}_B$ preserves the time-reversal symmetry. Note that even if both the Hamiltonian and the initial state are time-reversal invariant, the time-evolved wave function does not necessarily preserve this symmetry. Thus, the related dynamical symmetry requires a careful analysis of the final state, as shown in the Appendix.~\ref{App}.
The third transformation relation is given by
\begin{equation}\label{R}	    \mathcal{R}\mathcal{I}\hat{H}_B(\theta,\phi)\mathcal{I}^{-1}\mathcal{R}^{-1}=\hat{H}_B(-\theta,-\phi),
\end{equation}
which can be simply obtained from the transformation $\mathcal{T}\mathcal{K}\hat{H}_B\mathcal{K}^{\dagger}\mathcal{T}^{-1}$.

Based on the three transformations, we obtain the two following relations (see the Appendix~\ref{App}):
\begin{equation}\label{nsym1} \langle\hat{n}_{j,\sigma}(t)\rangle_{\theta}=\langle\hat{n}_{j',\sigma'}(t)\rangle_{-\theta},
\end{equation}
and 
\begin{equation}\label{nsym2} \langle\hat{n}_{j,\sigma}(t)\rangle_{U,\phi}=\langle\hat{n}_{j',\sigma'}(t)\rangle_{-U,-\phi},
\end{equation}
Here $j$ and $j'$ are two sites under reflection about the center of the chain, and $\sigma$ and $\sigma'$ denote opposite components. Notably, these two dynamical symmetry relations are exact, which are obtained via the symmetry analysis and verified by following numerical results from the exact diagonalization. They hold for a general class of initial states that preserve both inversion and time-reversal symmetries, without the constraints of particle number and product state. According to the two equations, two-component anyons flip their preferred expansion direction with opposite components when one changes the signs of $\{U,\phi\}$ or $\theta$ for the Hamiltonian $\hat{H}_A$ in Eq. (\ref{HamA}). These two equalities also imply that when $U=0$ or when $\{\theta,\phi\}=\{0,0\},\{0,\pi\},\{\pi,0\},\{\pi,\pi\}$, the expansion is symmetric, which is consistent with the numerical results shown in Fig.~\ref{fig2}. For the single-component anyon-Hubbard chain \cite{Alexey2018}, flipping the sign of the statistics phase $\theta$ or Hubbard interaction leads to the spatial inversion of the expansion. For two-component anyon-Hubbard chain considered here, by contrast, the change of $\theta$ sign gives rise to both the spatial inversion and the component flipping. Thus the spatial inversion of the expansion can be achieved by simultaneously flipping the statistics phase and two components, as described by Eq. (\ref{nsym1}). More interestingly, the dynamical symmetry described by Eq. (\ref{nsym2}) is unique to the two-component anyon-Hubbard chain with the synthetic flux, which indicates that the spatial inversion of the expansion can be achieved by simultaneously changing the signs of gauge flux and Hubbard interaction.

To show the two dynamical symmetries described by Eqs. \ref{nsym1} and \ref{nsym2}, we simulate the expansion dynamics from the two-particle initial state for various parameters in a system of size $L=26$, with the exact diagonalization results plotted in Fig. \ref{fig3}. In Fig.~\ref{fig3}(a), one can find $D^{(1)}_{\uparrow}(t)=-D^{(1)}_{\downarrow}(t)$ for any time $t$ when the sign of $\theta$ is reversed, which confirms the dynamical symmetry in Eq. (\ref{nsym1}). Fig.~\ref{fig3}(b) shows $D^{(1)}_{\uparrow}(t)=-D^{(1)}_{\downarrow}(t)$ when both the signs of $\phi$ and $U$ are reversed, indicating the dynamical symmetry in Eq. (\ref{nsym2}). 
Thus, the expansion dynamics of two-component anyons in this system are related by spatial inversion and component-flipping when we change the sign of $\theta$ or the signs of $\phi$ and $U$. In Figs.~\ref{fig3}(c) and \ref{fig3}(d), we plot $\Delta D^{(1)}=D^{(1)}_{\uparrow}-D^{(1)}_{\downarrow}$ at $t=4$ (in units of $\hbar/J$) in the $\theta$-$\phi$ and $U$-$\phi$ planes, respectively. 
The results show that $\Delta D^{(1)}(\theta,\phi)=\Delta D^{(1)}(-\theta,\phi)$ and 
$\Delta D^{(1)}(U,\phi)=\Delta D^{(1)}(-U,-\phi)$, as expected from the dynamical symmetries.

\section{\label{diff} Spreading suppression and chiral dynamics}

In this section, we further study the effects of the statistics phase  $\theta$ and synthetic gauge flux $\phi$ on the expansion dynamics of two anyons in the system. In the non-interacting case, we show the spreading is suppressed by increasing $\theta$ or $\phi$ from $0$ to $\pi$. In the interacting case with asymmetric transport, we find the chiral and antichiral dynamics that are tunable via $\theta$ and $\phi$.

\subsection{Spreading suppression for $U=0$}
\begin{figure}[t!]
	\centering
	\includegraphics[width=0.48\textwidth]{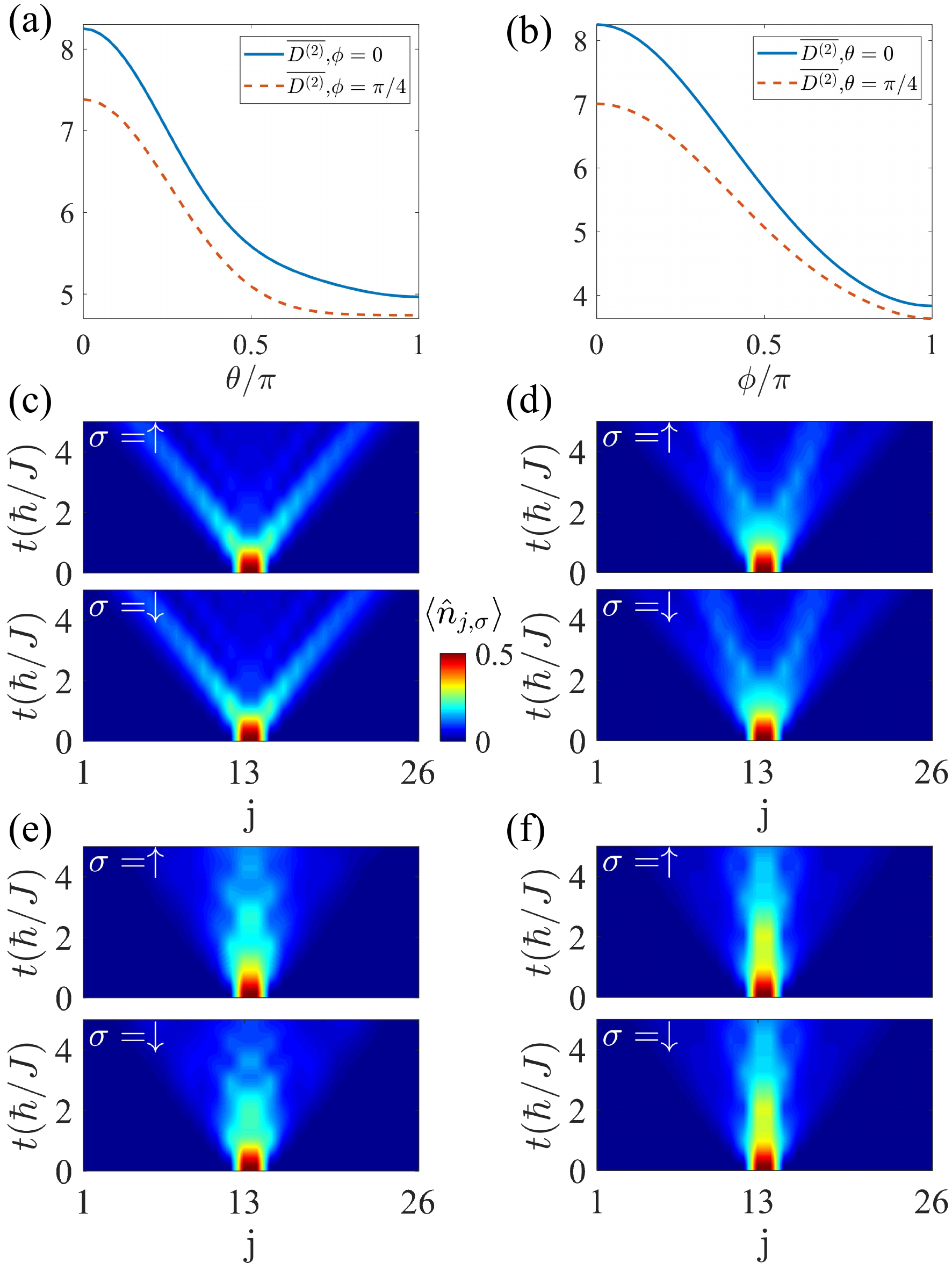}
	\caption{(Color online) $\overline{D^{(2)}}(t)$ at the time $t=2$ (in units of $\hbar/J$) as a function of (a) $\theta$ with $\phi=0,\pi/4$,  and (b) $\phi$ with $\theta=0,\pi/4$. Time evolution of density distributions for (c) $\theta=0$ and $\phi=0$; (d) $\theta=\pi/2$ and $\phi=0$; (e) $\theta=\pi/2$ and $\phi=\pi/2$; and (f) $\theta=\pi$ and $\phi=\pi/2$. Other parameters in (a-f) are $\Omega/J=1$, $U/J=0$, and $L=26$.}
	\label{fig4}
\end{figure}

We first consider the vanishing Hubbard interaction case with $U=0$, and show the exact diagonalization results of the two-anyon expansion dynamics in a system of size $L=26$ in Fig. \ref{fig4}. The overall spreading dynamics can be described by the second moment averaged over two components $\overline{D^{(2)}}(t)=[D^{(2)}_{\uparrow}(t) + D^{(2)}_{\downarrow}(t)]/2$. We plot the numerical results of $\overline{D^{(2)}}(t)$ at the time $t=2$ (in units of $\hbar/J$) as a function of $\theta$ for $\phi=0$ and $\pi/4$ in Fig. \ref{fig4}(a). For both $\phi=0$ and $\pi/4$, $\overline{D^{(2)}}$ decreases monotonically with increasing $\theta$ from $0$ to $\pi$, indicating that spreading is progressively suppressed as the statistics varies from bosonic to pseudofermionic. In addition, the values of $\overline{D^{(2)}}$ for $\phi=\pi/4$ are lower than those for $\phi=0$, which implies that a finite flux further suppresses the spreading. To see this point, we plot $\overline{D^{(2)}}$ (at the time $t=2$) as a function of $\phi$ for fixed $\theta=0$ and $\pi/4$ in Fig.~\ref{fig4}(b), which decreases monotonically as $\phi$ increases from $0$ to $\pi$. These results demonstrate the spreading suppression due to the statistics phase and gauge flux even in the case of $U=0$. To be more clearly, we show the time evolutions of two components for various values of $\theta$ and $\phi$ in Figs.~\ref{fig4}(c-f). For non-interacting bosons in the absence of synthetic gauge flux ($\theta=\phi=0$), the ballistic transport exhibits in the system~\cite{Ronzheimer2013}, as shown in Fig.~\ref{fig4}(c). For anyons with $\theta=\pi/2$ under a zero flux ($\phi=0$) in Fig.~\ref{fig4}(d), the spreading of both two components is suppressed, which exhibits the diffusive characteristics. This spreading suppression is further enhanced for anyons with $\theta=\pi/2$ under a non-zero flux $\phi=\pi/2$, as shown in Fig.~\ref{fig4}(e). For pseudofermions with $\theta=\pi$ under the flux $\phi=\pi/2$ in Fig.~\ref{fig4}(f), the spreading of two components becomes more localized.
For single-component anyons, anyonic statistics itself act as an effective interaction, thereby suppressing transport~\cite{Alexey2018}. In a bosonic ladder, the synthetic gauge flux also leads to transport suppression~\cite{Atala2014,Tai2017,Giri2023}. Our results show that the dynamical suppression can be induced by both the statistics phase and the gauge flux in the two-component anyon chain.

\begin{figure}[t!]
	\centering
	\includegraphics[width=0.48\textwidth]{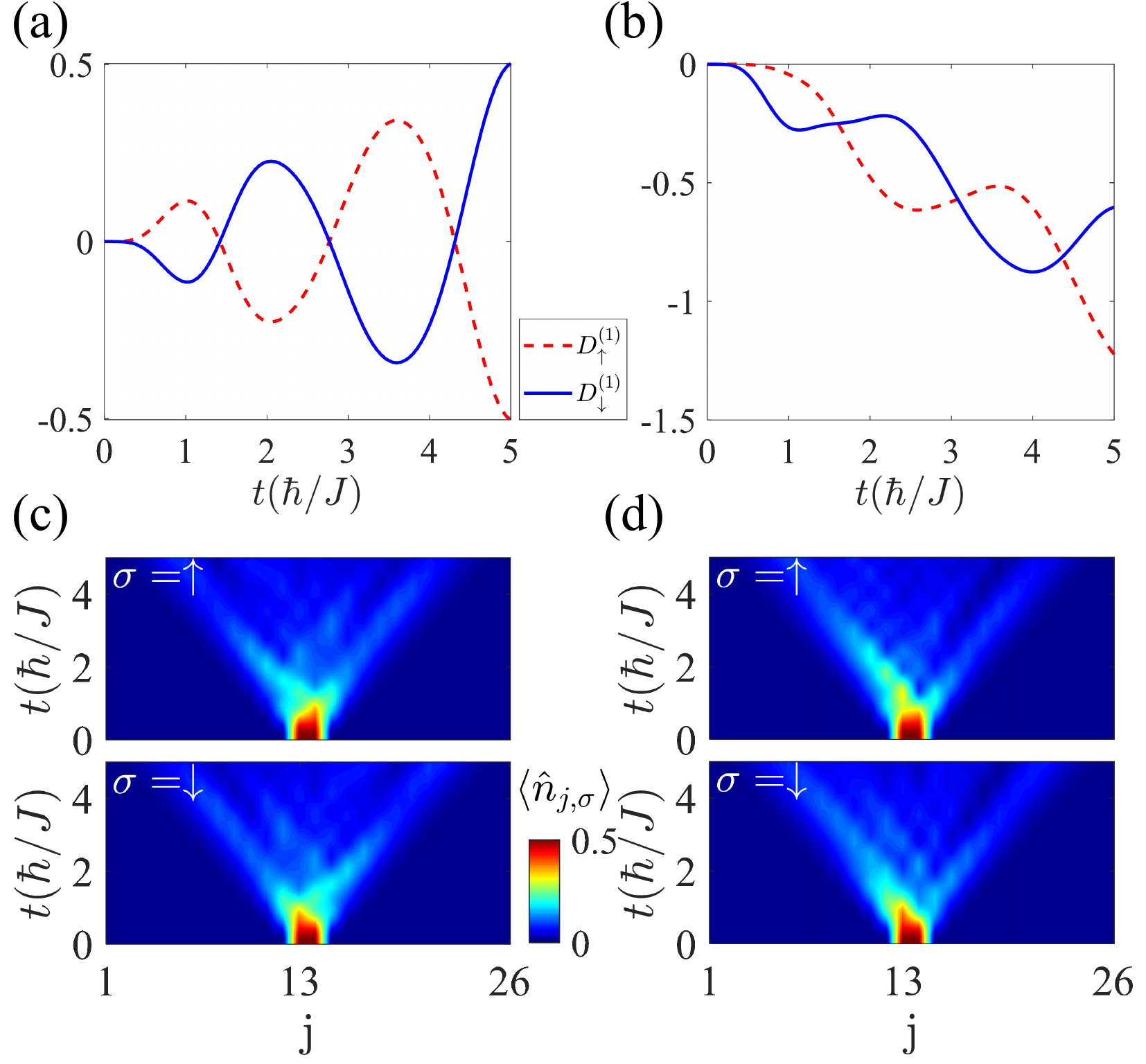}
	\caption{(Color online) Time evolution of $D^{(1)}_{\uparrow}$ and $D^{(1)}_{\downarrow}$ for (a) $\theta=0$ and (b) $\theta=\pi/2$. Time evolution of density distributions for (c) $\theta=0$ and (d) $\theta=\pi/2$. Other parameters in (a-d) are $U/J=4$, $\Omega/J=1$, $\phi=\pi/6$, and $L=26$.}
	\label{fig5}
\end{figure}
\begin{figure}[t!]
	\centering
	\includegraphics[width=0.45\textwidth]{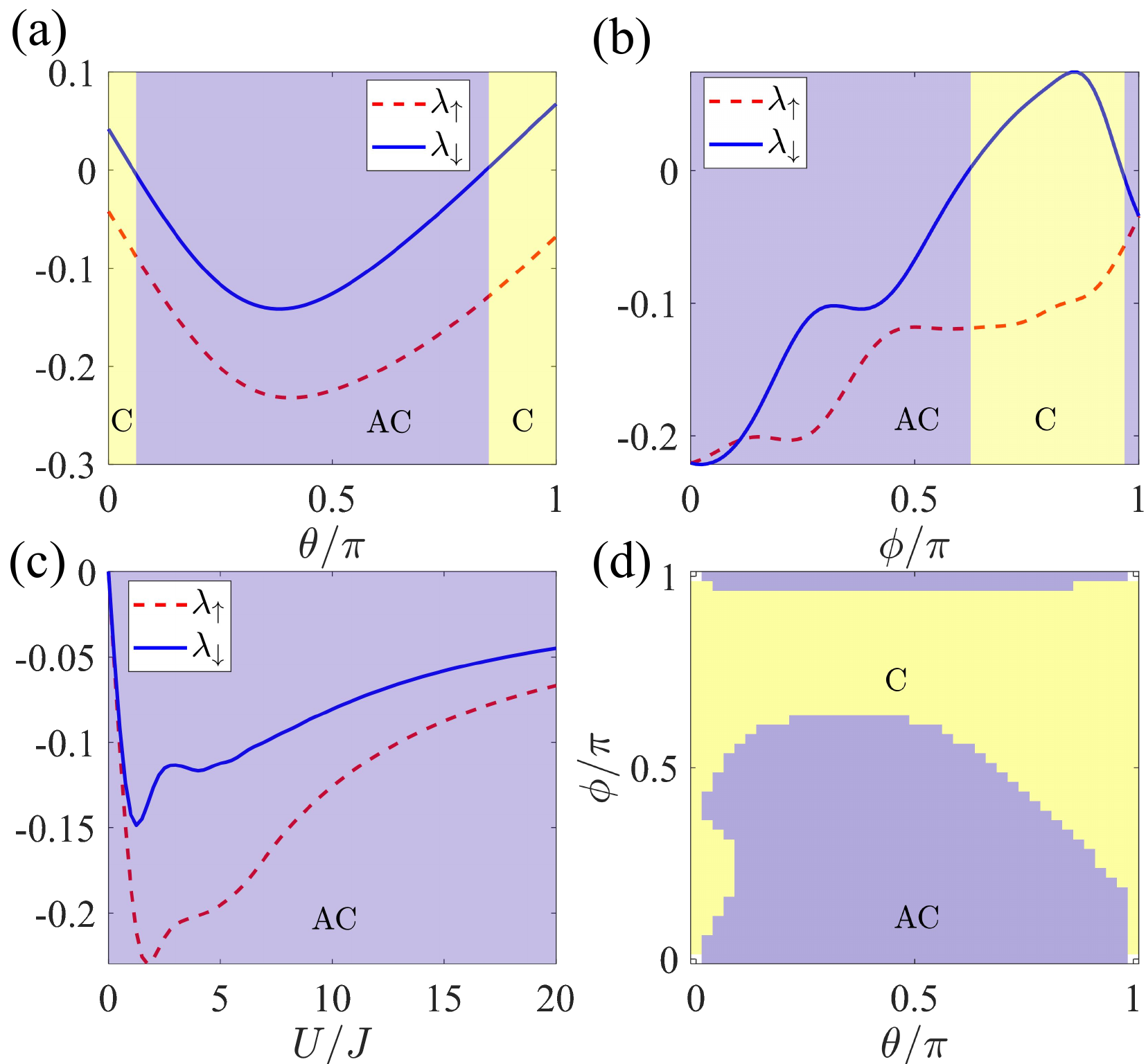}
	\caption{(Color online) $\lambda_{\uparrow}$ and $\lambda_{\downarrow}$ as a function of (a) $\theta/\pi$, (b) $\phi/\pi$, and (c) $U/J$. (d) Dynamical phase regimes in the $\theta$-$\phi$ plane determined by the sign of $\lambda_{\uparrow}\lambda_{\downarrow}$. Here, C denotes the chiral dynamics (yellow area) with $\lambda_{\uparrow}\lambda_{\downarrow}<0$, while AC denotes the antichiral dynamics (purple area) with $\lambda_{\uparrow}\lambda_{\downarrow}>0$. Other parameters are $\Omega/J=1$ and $L=26$ in (a-d), $U/J=4$ in (a,b,d), $\phi=\pi/4$ in (a), and $\theta=\pi/4$ in (b,c).}
	\label{fig6}
\end{figure}

\subsection{\label{c_ac}Chiral and antichiral dynamics for $U\neq 0$}

As previously discussed in Sec. \ref{sec_dysy}, the expansion of two-component anyons with finite Hubbard interaction is asymmetric in both spatial dimension and component degrees of freedom. It has been shown that the two-boson system ($\theta=0$) can exhibit chiral dynamics under the interplay between synthetic gauge flux and Hubbard interaction~\cite{Tai2017,Giri2023,Chen2024}. Here, we reveal that the asymmetric expansion of two anyons under the gauge flux leads to both the chiral and antichiral dynamics. The chiral and antichiral dynamics refer to the opposite and the same propagation directions for two components \cite{Wu2022,Chen2024,Ye2025}, respectively. The time evolutions of the center-of-mass of two components $D^{(1)}_{\sigma}$ generally oscillate with drifts, such as those of opposite or same directions shown in Fig. \ref{fig5}(a) and (b). The propagation direction of each component can be extracted from its center-of-mass drift. Figs. \ref{fig5}(c) and (d) show the examples of density evolution for chiral and antichiral dynamics, respectively. To this end, we numerically extract the propagation direction by linearly fitting the time evolution of the center-of-mass as  $D^{(1)}_{\sigma}(t)=\lambda_{\sigma}t$ from $t=0$ to $t=5$ (in units of $\hbar/J$), where $\lambda_{\sigma}$ is the fitting slope. The sign of $\lambda_{\sigma}$ denotes the propagation of $\sigma$-component along the positive or negative direction. Thus, chiral and antichiral dynamics exhibit when $\lambda_{\uparrow}\lambda_{\downarrow}<0$ and $\lambda_{\uparrow}\lambda_{\downarrow}>0$, respectively.

Figures~\ref{fig6}(a-c) show the numerical results of $\lambda_{\uparrow}$ and $\lambda_{\downarrow}$ as functions of $\theta$, $\phi$, and $U$, respectively. From the sign of $\lambda_{\uparrow}\lambda_{\downarrow}$, we can identify the chiral (yellow shading) and antichiral (purple shading) dynamics with respect to these paramters. For bosons (pseudofermions) with $\theta=0$ ($\theta=\pi$), the chiral dynamics exhibits under finite flux and Hubbard interaction. By increasing the statistics phase $\theta$, the chiral dynamics of anyons can becomes antichial and then returns back, as shown in Fig. \ref{fig6}(a). For anyons, the chiral and antichiral dynamics is also tunable via $\phi$, as shown in Fig. \ref{fig6}(b). This chiral-antichiral crossover is due to the interplay between the chiral transport under gauge flux and the asymmetric transport of anyons with finite $U$. For fixed values of $\theta$ and $\phi$, increasing $U$ does not change the signs of $\lambda_{\uparrow}$ and $\lambda_{\downarrow}$, but reduces their magnitudes, as shown in Fig. \ref{fig6}(c). This demonstrates that varying $U$ does not induce the changing between the chiral and antichiral dynamics. Instead, it suppresses the overall propagation in both two dynamics. Finally, we numerically obtain the dynamical phase regimes in the $\theta$-$\phi$ plane, as shown in Fig.~\ref{fig6}(d). Note that in Fig.~\ref{fig6}(d), the parameter points for $\phi=\{0,\pi\}$ and $\theta=\{0,\pi\}$ are left blank. The chiral and antichiral dynamics are not well defined at these specific points, as the characteristic directional bias and component-dependence are absent under the time-reversal and inversion symmetries therein.

The physical origin of the chiral-antichiral crossover is rooted in the competition between the asymmetric transport and the chiral motion. In the bosonic limit with $\theta=0$, the magnetic flux drives the two components of interacting bosons to propagate in opposite directions, which leads to the chiral dynamics. In the absence of the flux with $\phi=0$, the two components of particles share the same nontrivial statistics phase, which breaks the inversion symmetry and drives both components to propagate along the same direction. In the parameter region of finite $\theta$ and $\phi$, the competition between co-directional and chiral motions is expected to gives rise to the crossover between the chiral and antichiral dynamics.

\begin{figure}[t!]
	\centering
	\includegraphics[width=0.48\textwidth]{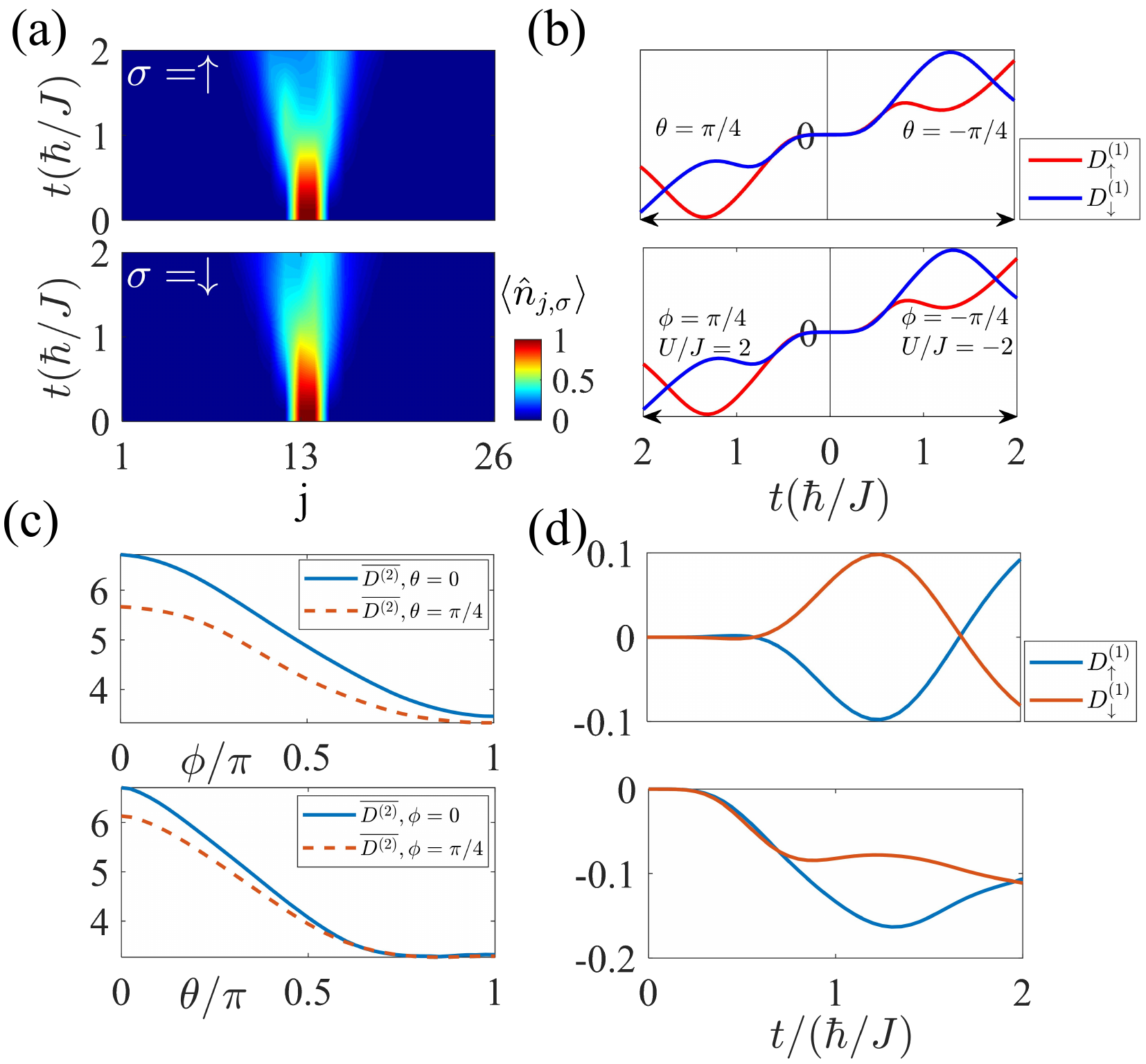}
	\caption{(Color online) Four-particle dynamics. (a) Time evolution of density distributions for $U/J=0.5$, $\theta=\pi/5$ and $\phi=\pi/5$. (b) Time evolution of $D^{(1)}_{\uparrow}$ and $D^{(1)}_{\downarrow}$ for (top) $\theta=\pm\pi/4$ and (bottom) $\phi=\pm\pi/4$ and $U/J=\pm 2$. (c) $\overline{D^{(2)}}(t)$ at the time $t=2$ (in units of $\hbar/J$) as a function of (top) $\phi=0,\pi/4$ and (bottom) $\theta=0,\pi/4$. (d) Time evolution of $D^{(1)}_{\uparrow}$ and $D^{(1)}_{\downarrow}$ for (top) $\theta=0$ and (bottom) $\pi/2$. Other parameters are $\Omega/J=1$ and $L=26$ in (a-d), $U/J=2$ and $\phi=\pi/4$ in the top of (b), $\theta=\pi/4$ in the bottom of (b), $U/J=0$ in (c) and $U/J=2$ and $\phi=\pi/2$ in (d).}\label{fig7}

\end{figure}

\section{\label{conclusion}Discussion and Conclusion}

We note the dynamical properties revealed above preserve for other initial states satisfying inversion and time-reversal symmetries, as indicated from the symmetry analysis. To further demonstrate the generality of our main results for more particles, we consider a four-particle initial state $|\psi_0\rangle=\hat{a}_{j_0,\uparrow}^{\dagger}\hat{a}_{j_0,\downarrow}^{\dagger}\hat{a}_{j_0+1,\uparrow}^{\dagger}\hat{a}_{j_0+1,\downarrow}^{\dagger}|\text{vac}\rangle$, which is occupied symmetrically around the lattice center and preserves the time-reversal symmetry. We simulate the four-particle dynamics using the time-dependent variational principle algorithm \cite{Fishman2022}, with typical results shown in Fig.~\ref{fig7}. The asymmetric and component-imbalance transports of the two-component anyons in this four-particle system under non-zero synthetic flux $\phi$ and statistics angle $\theta$ are shown in Fig.~\ref{fig7}(a), similar to the two-particle case. As shown in Fig.~\ref{fig7}(b), the four-particle dynamics also are symmetric with respect to the system parameters $\{\theta, \phi, U\}$, described by Eqs. (\ref{nsym1}) and (\ref{nsym2}). In Fig.~\ref{fig7}(c), the overall trend of $\overline{D^{(2)}}$ as a function of $\theta$ and $\phi$ is consistent with that in Figs.~\ref{fig4}(a) and \ref{fig4}(b). This indicates that the dynamics in the four-particle case are suppressed by increasing $\theta$ or $\phi$ from 0 to $\pi$. We further demonstrate the chiral and antichiral dynamics in the interacting four-particle case. As shown in Fig.~\ref{fig7}(d), the center-of-mass of two components propagate in opposite directions with non-zero flux and zero statistics angle, while for a non-zero statistics angle, the two components propagate along the same direction. Thus, in the four-particle case, the chiral and antichiral dynamics can be tuned by the statistics angle and gauge flux.

We note that these results can survive beyond the few-particle simulations. The asymmetric transport arising from breaking the exact dynamical symmetries in Eqs. (12) and (13) are independent of the particle number of an initial state. The chiral dynamics arising from the synthetic flux can also exhibit in the many-particle case. For instance, the flux-induced chiral dynamics in the many-body bosonic ladder with $\theta=0$ have been shown for the quarter-filling and half-filling cases \cite{Piraud2015}. For our anyonic ladder in the case of larger particle numbers, one can expect that the statistical phase generally gives rise to spatially asymmetric transport and its interplay with the flux-induced chiral dynamics naturally leads to the crossover from chiral to antichiral dynamics.

In summary, we have explored the non-equilibrium dynamics in a one-dimensional two-component anyon-Hubbard model with synthetic gauge flux that would be realizable with ultracold bosonic atoms. We have revealed the asymmetric and imbalance expansions and the dynamical symmetries under the statistics phase and synthetic gauge flux in this system. We have further demonstrated that both the statistics phase and gauge flux suppress the spreading, and induce tunable chiral and antichiral dynamics within different parameter regimes. These findings highlight the rich dynamical phenomena of multi-component anyons arising from the interplay of anyonic statistics, synthetic gauge fields, and interactions. Our work also provides a symmetry-based framework for further studying multi-component anyons. Since the density-dependent phase for effective anyonic statistics phase and the synthetic gauge fluxes in two-leg ladders have been experimentally realized with tunable bosonic atoms in optical lattices \cite{Greiner2024,Atala2014,Tai2017}, we expect that the two-component anyon-Hubbard chain will be realized to explore these novel non-equilibrium dynamics in the future experiments.

\appendix
\section{Derivations of dynamical symmetries} \label{App}

Here we present some details of analyzing and deriving the dynamical symmetries. We consider a generic initial state $|\psi_0\rangle$ preserving both inversion and time-reversal symmetry, which can be simply a product state in Fock space or even an entangled state. We first consider two target states which are related by time-reversal symmetry $|\psi_1\rangle=\mathcal{T}|\psi_2\rangle$.
Using a Taylor expansion, the time evolution operator can be written as
\begin{equation} \mathcal{U}=e^{-i\hat{H}_Bt}=\sum_{n=0}^{\infty}\frac{(-i\hat{H}_Bt)^n}{n!}.
\end{equation}
The matrix element corresponding to the $k$-th order term that evolves $|\psi_0\rangle$ to $|\psi_1\rangle$ is
\begin{equation}
	M_k^{(1)}=\langle\psi_1|\frac{(-i\hat{H}_B t)^k}{k!}|\psi_0\rangle=\frac{(-it)^k}{k!}\langle\psi_1|\hat{H}_B^k(\theta)|\psi_0\rangle.
\end{equation}
Similarly we define
\begin{equation}
	\begin{split}
     M_k^{(2)}&=\langle\psi_2|\frac{(-i\hat{H}_B t)^k}{k!}|\psi_0\rangle\\&=\langle\psi_1|\mathcal{T}\frac{(-i\hat{H}_B t)^k}{k!}\mathcal{T}^{-1}|\psi_0\rangle\\&=\frac{(-it)^k}{k!}\langle\psi_1|(-1)^k\hat{H}_B^k(-\theta)|\psi_0\rangle.
	\end{split}
\end{equation}

We take $k$ to be the lowest order in the perturbation expansion for which both $M_k^{(1)}$ and $M_k^{(2)}$ are non-zero. The $(k+1)$-th order also contributes to the evolution because of the on-site interaction $U$, since the interaction term does not change the state configuration. All higher orders act similarly to the $(k+1)$-th order, as the interaction term keeps the wavefunction in the same configuration once it is reached at the $k$-th order. Thus, even though we present only the $k$-th and $(k+1)$-th orders and treat higher orders as perturbative contributions in the following, our analytical derivations are still exact. We define $S_1$ and $S_2$ to be the amplitudes including the contribution of the leading terms $M_k^{(1,2)}$ and $M_{k+1}^{(1,2)}$ as
\begin{equation}
	\begin{split}			
		S_1&=|M_k^{(1)}+M_{k+1}^{(1)}|+\mathcal{O}_1\\&\approx\frac{t^k}{k!}|\langle\psi_1|\hat{H}_B^k(\theta)|\psi_0\rangle-\frac{it}{k+1}\langle\psi_1|\hat{H}_B^{k+1}(\theta)|\psi_0\rangle|,
	\end{split}
\end{equation} 
and
\begin{equation}
	\begin{split}			
		S_2&=|M_k^{(2)}+M_{k+1}^{(2)}|+\mathcal{O}_2\\&\approx\frac{t^k}{k!}|\langle\psi_1|\hat{H}_B^k(-\theta)|\psi_0\rangle+\frac{it}{k+1}\langle\psi_1|\hat{H}_B^{k+1}(-\theta)|\psi_0\rangle|.
	\end{split}
\end{equation}
Here $\mathcal{O}_{(1,2)}$ are the pertubative terms.

For vanishing Hubbard interaction with $U=0$, the matrix element corresponding to the $(k+1)$-th order term vanishes, since hopping or inter-component coupling once more could not return to the same configuration as the target states. In this case, $S_1=S_2$ when either $\theta$ or $\phi$ equals to $0$ or $\pi$. $S_1$ and $S_2$ are not necessarily equal to each other unless both $\theta$ and $\phi$ are either 0 or $\pi$. This implies that for fractional statistics angle $\theta$ and gauge flux $\phi$, the perturbation analysis predicts asymmetric density expansion, indicating the breaking of time-reversal symmetry when $U=0$. For $U\neq0$, $S_1$ and $S_2$ are generally unequal for arbitrary $\theta$. However, for $\phi=0$ or $\pi$, $S_1$ and $S_2$ are equal. This is because $\hat{H}_B(-\theta)=\hat{H}^*_B(\theta)$ when $\phi=0$ or $\pi$. Thus,
\begin{equation}
	\begin{split}		 	S_2&=\frac{t^k}{k!}|\langle\psi_1|[\hat{H}_B^k(-\theta)]^*|\psi_0\rangle+\frac{it}{k+1}\langle\psi_1|[\hat{H}_B^{k+1}(-\theta)]^*|\psi_0\rangle|\\&=\frac{t^k}{k!}|\langle\psi_1|\hat{H}_B^k(\theta)|\psi_0\rangle-\frac{it}{k+1}\langle\psi_1|\hat{H}_B^{k+1}(\theta)|\psi_0\rangle|\\&=S_1.
	\end{split}	
\end{equation}
Therefore, the dynamics under finite Hubbard interaction are time-reversal symmetric when $\phi= 0$ or $\pi$.

Next, we consider another pair of target states which are related by the inversion symmetry, $|\psi_3\rangle=\mathcal{I}|\psi_1\rangle$. Similarly, we can obtain the following equations:
\begin{equation}
	\begin{split}
		M_k^{(3)}&=\langle\psi_3|\frac{[-i\hat{H}_B(\theta,\phi) t]^k}{k!}|\psi_0\rangle\\&=\langle\psi_1|\mathcal{I}\frac{[-i\hat{H}_B(\theta,\phi) t]^k}{k!}\mathcal{I}^{\dagger}|\psi_0\rangle\\&=\langle\psi_1|\mathcal{I}\mathcal{K}^{\dagger}\frac{[-i\hat{H}_B(\theta,-\phi) t]^k}{k!}\mathcal{K}\mathcal{I}^{\dagger}|\psi_0\rangle\\&=e^{i(\xi_3-\xi_0)}\langle\psi_1|\frac{[-i\hat{H}_B(-\theta,-\phi) t]^k}{k!}|\psi_0\rangle,
	\end{split}
\end{equation}
where the global phase factor $e^{i(\xi_3-\xi_0)}$ is associated with the action of $\mathcal{R}=e^ {i\theta[\sum_{j,\sigma}^{L}\hat{n}_{j,\sigma}(\hat{n}_{j,\sigma}-1)/2+\sum_{j}^{L}\hat{n}_{j,\uparrow}\hat{n}_{j,\downarrow}]}$ operator and $\mathcal{K}=\mathcal{R}\mathcal{I}\mathcal{T}$.
Then the amplitude including the total contribution
of the $k$-th and the $(k+1)$-th orders is
\begin{equation}
	\begin{split}			&S_3=|M_k^{(3)}+M_{k+1}^{(3)}|+\mathcal{O}_3\\&\approx\frac{t^k}{k!}|\langle\psi_1|[\hat{H}_B^k(-\theta,-\phi)-\frac{it}{k+1}\hat{H}_B^{k+1}(-\theta,-\phi)]|\psi_0\rangle|\\&=\frac{t^k}{k!}|\langle\psi_1|\hat{H}_B^k(\theta,\phi)|\psi_0\rangle+\frac{it}{k+1}\langle\psi_1|\hat{H}_B^{k+1}(\theta,\phi)|\psi_0\rangle|.
	\end{split}
\end{equation}
where $\mathcal{O}_3$ is the pertubative terms.

For $U=0$, $S_1$ and $S_3$ are the same due to the vanishing of the $(k+1)$-th order term. This implies that inversion symmetry is preserved in the density expansion. For $U\neq0$, if both $\theta$ and $\phi$ are 0 or $\pi$, the matrix elements $\langle\psi_1|\hat{H}_B^k(\theta,\phi)|\psi_0\rangle$ and $\langle\psi_1|\hat{H}_B^{k+1}(\theta,\phi)|\psi_0\rangle$ are real numbers. In this case, $S_1$ is also equal to $S_3$. Conversely, for fractional statistics angle $\theta$ and gauge flux $\phi$, these two matrix elements are generally complex numbers. Thus, $S_1$ and $S_3$ are not necessarily equal which implies the breaking of inversion symmetry in the density expansion.

Moreover we have the following dynamical symmetries
\begin{equation}\label{sym1}
	\begin{split}		   &\langle\hat{n}_{j,\sigma}(t)\rangle_{\phi}=\langle\psi_0|e^{i\hat{H}_B(\phi)t}\hat{n}_{j,\sigma}e^{-i\hat{H}_B(\phi)t}|\psi_0\rangle\\&=\langle\psi_0|\mathcal{K}^{\dagger}e^{-i\hat{H}_B(-\phi)t}\mathcal{K}\hat{n}_{j,\sigma}\mathcal{K}^{\dagger}e^{i\hat{H}_B(-\phi)t}\mathcal{K}|\psi_0\rangle\\&=\langle\psi_0|\mathcal{T}e^{-i\hat{H}_B(-\phi)t}\mathcal{I}\hat{n}_{j,\sigma'}\mathcal{I}^{\dagger}e^{i\hat{H}_B(-\phi)t}\mathcal{T}^{-1}|\psi_0\rangle\\&=\langle\psi_0|\mathcal{T}e^{-i\hat{H}_B(-\phi)t}\hat{n}_{j',\sigma'}e^{i\hat{H}_B(-\phi)t}\mathcal{T}^{-1}|\psi_0\rangle,
	\end{split}
\end{equation}
where $\sigma$ and $\sigma'$ denote opposite components, $j$ and $j'$ are two sites related by reflection about the chain center.
Using the time-reversal relation $\mathcal{T}\hat{H}_B(\theta)\mathcal{T}^{-1}=\hat{H}_B(-\theta)$, we obtain
\begin{equation}
	\begin{split}		&\langle\hat{n}_{j,\sigma}(t)\rangle_{\theta,\phi}=\langle\psi_0|\mathcal{T}e^{-i\hat{H}_B(\theta,-\phi)t}\hat{n}_{j',\sigma'}e^{i\hat{H}_B(\theta,-\phi)t}\mathcal{T}^{-1}|\psi_0\rangle\\&=\langle\psi_0|e^{i\hat{H}_B(-\theta,-\phi)t}\mathcal{T}^{-1}\hat{n}_{j',\sigma'}\mathcal{T}e^{i\hat{H}_B(-\theta,-\phi)t}|\psi_0\rangle\\&=\langle\psi_0|\Gamma e^{i\hat{H}_B(-\theta,\phi)t}\Gamma^{\dagger}\mathcal{T}^{-1}\hat{n}_{j',\sigma'}\mathcal{T}\Gamma e^{i\hat{H}_B(-\theta,\phi)t}\Gamma^{\dagger}|\psi_0\rangle\\&=\langle\hat{n}_{j',\sigma'}(t)\rangle_{-\theta,\phi}.
	\end{split}
\end{equation}
Here $\Gamma$ is the component-flipping operator, i.e., $\Gamma\sigma\Gamma^{\dagger}=\sigma'$. This relation implies that when we change the sign of $\theta$, the density expansions are related by inversion and component-flipping. In addition, we define a number parity operator $\mathcal{P}=e^{i\pi[\sum_{r}^{L/2}(\hat{n}_{2r+1,\sigma}+\hat{n}_{2r+1,\sigma'})+\sum_{j}^{L}\hat{n}_{j,\sigma}]}$. This operator anti-commutes with the hopping term and the inter-component coupling term, but commutes with the on-site interaction terms. Therefore, we obtain
\begin{equation}
	\mathcal{P}\hat{H}_B(U)\mathcal{P}^{\dagger}=-\hat{H}_B(-U).
\end{equation}
Then we have
\begin{equation}\label{sym2}
	\begin{split}
		&\langle\hat{n}_{j,\sigma}(t)\rangle_{\phi,U}=\langle\psi_0|e^{i\hat{H}_B(\phi,U)t}\hat{n}_{j,\sigma}e^{-i\hat{H}_B(\phi,U)t}|\psi_0\rangle\\&=\langle\psi_0|\mathcal{K}^{\dagger}e^{-i\hat{H}_B(-\phi,U)t}\mathcal{K}\hat{n}_{j,\sigma}\mathcal{K}^{\dagger}e^{i\hat{H}_B(-\phi,U)t}\mathcal{K}|\psi_0\rangle\\&=\langle\psi_0|e^{-i\hat{H}_B(-\phi,U)t}\mathcal{I}\hat{n}_{j,\sigma'}\mathcal{I}^{\dagger}e^{i\hat{H}_B(-\phi,U)t}|\psi_0\rangle\\&=\langle\psi_0|e^{-i\hat{H}_B(-\phi,U)t}\hat{n}_{j',\sigma'}e^{i\hat{H}_B(-\phi,U)t}|\psi_0\rangle\\&=\langle\psi_0|\mathcal{P}^{-1}e^{i\hat{H}_B(-\phi,-U)t}\mathcal{P}\hat{n}_{j',\sigma'}\mathcal{P}^{-1}e^{-i\hat{H}_B(-\phi,-U)t}\mathcal{P}|\psi_0\rangle\\&=\langle\psi_0|e^{i\hat{H}_B(-\phi,-U)t}\hat{n}_{j',\sigma'}e^{-i\hat{H}_B(-\phi,-U)t}|\psi_0\rangle\\&=\langle\hat{n}_{j',\sigma'}(t)\rangle_{-\phi,-U}.
	\end{split}
\end{equation}
Thus, the reversing the signs of $\phi$ and $U$ implies that the expansions are related by inversion and component exchange.

\begin{acknowledgments}
This work is supported by the Guangdong Basic and Applied Basic Research Foundation (Grant No. 2024B1515020018), the National Natural Science Foundation of China (Grant No. 12174126), the Guangdong Provincial Quantum Science Strategic Initiative (Grants No. GDZX2204003 and No. GDZX2404001), the Science and Technology Program of Guangzhou (Grant No. 2024A04J3004), and the Open Fund of Key Laboratory of Atomic and Subatomic Structure and Quantum Control (Ministry of Education). 
\end{acknowledgments}

\bibliography{reference}

@article{Leinaas1977,
	author = {Leinaas, J. M. and Myrheim, J.},
	title = {On the theory of identical particles},
	journal = {Nuovo Cimento Soc. Ital. Fis. B},
	volume = {37},
	pages = {1-23},
	ISSN = {1826-9877},
	DOI = {10.1007/BF02727953},
	url = {https://doi.org/10.1007/BF02727953},
	year = {1977},
}

@article{Wilczek1982,
	title = {Magnetic Flux, Angular Momentum, and Statistics},
	author = {Wilczek, Frank},
	journal = {Phys. Rev. Lett.},
	volume = {48},
	issue = {17},
	pages = {1144--1146},
	numpages = {0},
	year = {1982},
	month = {Apr},
	publisher = {American Physical Society},
	doi = {10.1103/PhysRevLett.48.1144},
	url = {https://link.aps.org/doi/10.1103/PhysRevLett.48.1144}
}

@article{Haldane1991,
	title = {``Fractional statistics'' in arbitrary dimensions: A generalization of the Pauli principle},
	author = {Haldane, F. D. M.},
	journal = {Phys. Rev. Lett.},
	volume = {67},
	issue = {8},
	pages = {937--940},
	numpages = {0},
	year = {1991},
	month = {Aug},
	publisher = {American Physical Society},
	doi = {10.1103/PhysRevLett.67.937},
	url = {https://link.aps.org/doi/10.1103/PhysRevLett.67.937}
}

@article{Wilczek2024,
	author = {Greiter, Martin and Wilczek, Frank},
	title = {Fractional Statistics}, 
	journal= {Annu. Rev. Condens. Matter Phys},
	year = {2024},
	volume = {15},
	pages = {131-157},
	doi = {10.1146/annurev-conmatphys-040423-014045},
	url = {https://doi.org/10.1146/annurev-conmatphys-040423-014045},
	publisher = "Annual Reviews",
	issn = {1947-5462},
	type = {Journal Article},
}

@article{Laughlin1983,
	title = {Anomalous Quantum Hall Effect: An Incompressible Quantum Fluid with Fractionally Charged Excitations},
	author = {Laughlin, R. B.},
	journal = {Phys. Rev. Lett.},
	volume = {50},
	issue = {18},
	pages = {1395--1398},
	numpages = {0},
	year = {1983},
	month = {May},
	publisher = {American Physical Society},
	doi = {10.1103/PhysRevLett.50.1395},
	url = {https://link.aps.org/doi/10.1103/PhysRevLett.50.1395}
}

@article{Halperin1984,
	title = {Statistics of Quasiparticles and the Hierarchy of Fractional Quantized Hall States},
	author = {Halperin, B. I.},
	journal = {Phys. Rev. Lett.},
	volume = {52},
	issue = {18},
	pages = {1583--1586},
	numpages = {0},
	year = {1984},
	month = {Apr},
	publisher = {American Physical Society},
	doi = {10.1103/PhysRevLett.52.1583},
	url = {https://link.aps.org/doi/10.1103/PhysRevLett.52.1583}
}

@article{Kim2005,
	title = {Signatures of Fractional Statistics in Noise Experiments in Quantum Hall Fluids},
	author = {Kim, Eun-Ah and Lawler, Michael and Vishveshwara, Smitha and Fradkin, Eduardo},
	journal = {Phys. Rev. Lett.},
	volume = {95},
	issue = {17},
	pages = {176402},
	numpages = {4},
	year = {2005},
	month = {Oct},
	publisher = {American Physical Society},
	doi = {10.1103/PhysRevLett.95.176402},
	url = {https://link.aps.org/doi/10.1103/PhysRevLett.95.176402}
}

@article{Coldea2001,
	title = {Experimental Realization of a 2D Fractional Quantum Spin Liquid},
	author = {Coldea, R. and Tennant, D. A. and Tsvelik, A. M. and Tylczynski, Z.},
	journal = {Phys. Rev. Lett.},
	volume = {86},
	issue = {7},
	pages = {1335--1338},
	numpages = {0},
	year = {2001},
	month = {Feb},
	publisher = {American Physical Society},
	doi = {10.1103/PhysRevLett.86.1335},
	url = {https://link.aps.org/doi/10.1103/PhysRevLett.86.1335}
}

@article{Kitaev2006,
	author = {Alexei Kitaev},
	title = {Anyons in an exactly solved model and beyond},
	journal = {Annals of Physics},
	volume = {321},
	number = {1},
	pages = {2-111},
	year = {2006},
	note = {January Special Issue},
	issn = {0003-4916},
	doi = {https://doi.org/10.1016/j.aop.2005.10.005},
	url = {https://www.sciencedirect.com/science/article/pii/S0003491605002381},
}

@article{Yao2007,
	title = {Exact Chiral Spin Liquid with Non-Abelian Anyons},
	author = {Yao, Hong and Kivelson, Steven A.},
	journal = {Phys. Rev. Lett.},
	volume = {99},
	issue = {24},
	pages = {247203},
	numpages = {4},
	year = {2007},
	month = {Dec},
	publisher = {American Physical Society},
	doi = {10.1103/PhysRevLett.99.247203},
	url = {https://link.aps.org/doi/10.1103/PhysRevLett.99.247203}
}

@article{Bauer2014,
	author = {Bauer, B. and Cincio, L. and Keller, B. P. and Dolfi, M. and Vidal, G. and Trebst, S. and Ludwig, A. W. W.},
	title = {Chiral spin liquid and emergent anyons in a Kagome lattice Mott insulator},
	journal = {Nat. Commun.},
	volume = {5},
	pages = {5137},
	ISSN = {2041-1723},
	DOI = {10.1038/ncomms6137},
	url = {https://doi.org/10.1038/ncomms6137},
	year = {2014},
	type = {Journal Article}
}

@Article{Semeghini2021,
  author  = {G. Semeghini and H. Levine and A. Keesling and S. Ebadi and T. T. Wang and D. Bluvstein and R. Verresen and H. Pichler and M. Kalinowski and R. Samajdar and A. Omran and S. Sachdev and A. Vishwanath and M. Greiner and V. Vuletić and M. D. Lukin},
  journal = {Science},
  title   = {Probing topological spin liquids on a programmable quantum simulator},
  year    = {2021},
  number  = {6572},
  pages   = {1242-1247},
  volume  = {374},
  doi     = {10.1126/science.abi8794},
  url     = {https://doi/abs/10.1126/science.abi8794},
}

@article{Kitaev2003,
	author = {A.Yu. Kitaev},
	title = {Fault-tolerant quantum computation by anyons},
	journal = {Annals of Physics},
	volume = {303},
	number = {1},
	pages = {2-30},
	year = {2003},
	issn = {0003-4916},
	doi = {doi.org/10.1016/S0003-4916(02)00018-0},
	url = {https://doi.org/10.1016/S0003-4916(02)00018-0},
}

@article{Sarma2005,
	title = {Topologically Protected Qubits from a Possible Non-Abelian Fractional Quantum Hall State},
	author = {Das Sarma, Sankar and Freedman, Michael and Nayak, Chetan},
	journal = {Phys. Rev. Lett.},
	volume = {94},
	issue = {16},
	pages = {166802},
	numpages = {4},
	year = {2005},
	month = {Apr},
	publisher = {American Physical Society},
	doi = {10.1103/PhysRevLett.94.166802},
	url = {https://link.aps.org/doi/10.1103/PhysRevLett.94.166802}
}

@Article{Lindner2013,
  author  = {Ady Stern and Netanel H. Lindner},
  journal = {Science},
  title   = {Topological Quantum Computation—From Basic Concepts to First Experiments},
  year    = {2013},
  number  = {6124},
  pages   = {1179-1184},
  volume  = {339},
  doi     = {10.1126/science.1231473},
  url     = {https://10.1126/science.1231473},
}

@article{Google2023,
	author = {Google Quantum AI and Collaborators},
	title = {Non-Abelian braiding of graph vertices in a superconducting processor},
	journal = {Nature},
	volume = {618},
	number = {7964},
	pages = {264-269},
	ISSN = {1476-4687},
	DOI = {10.1038/s41586-023-05954-4},
	url = {https://10.1038/s41586-023-05954-4},
	year = {2023},
	type = {Journal Article}
}

@article{Ha1994,
	title = {Exact Dynamical Correlation Functions of Calogero-Sutherland Model and One-Dimensional Fractional Statistics},
	author = {Ha, Z. N. C.},
	journal = {Phys. Rev. Lett.},
	volume = {73},
	issue = {12},
	pages = {1574--1577},
	numpages = {0},
	year = {1994},
	month = {Sep},
	publisher = {American Physical Society},
	doi = {10.1103/PhysRevLett.73.1574},
	url = {https://link.aps.org/doi/10.1103/PhysRevLett.73.1574}
}

@article{Murthy1994,
	title = {Thermodynamics of a One-Dimensional Ideal Gas with Fractional Exclusion Statistics},
	author = {Murthy, M. V. N. and Shankar, R.},
	journal = {Phys. Rev. Lett.},
	volume = {73},
	issue = {25},
	pages = {3331--3334},
	numpages = {0},
	year = {1994},
	month = {Dec},
	publisher = {American Physical Society},
	doi = {10.1103/PhysRevLett.73.3331},
	url = {https://link.aps.org/doi/10.1103/PhysRevLett.73.3331}
}

@article{Yue1995,
	title = {Bosonization of One-Dimensional Exclusons and Characterization of Luttinger Liquids},
	author = {Wu, Yong-Shi and Yu, Yue},
	journal = {Phys. Rev. Lett.},
	volume = {75},
	issue = {5},
	pages = {890--893},
	numpages = {0},
	year = {1995},
	month = {Jul},
	publisher = {American Physical Society},
	doi = {10.1103/PhysRevLett.75.890},
	url = {https://link.aps.org/doi/10.1103/PhysRevLett.75.890}
}

@article{Amico1998,
	title = {One-dimensional $\mathrm{XXZ}$ model for particles obeying fractional statistics},
	author = {Amico, Luigi and Osterloh, Andreas and Eckern, Ulrich},
	journal = {Phys. Rev. B},
	volume = {58},
	issue = {4},
	pages = {R1703--R1706},
	numpages = {0},
	year = {1998},
	month = {Jul},
	publisher = {American Physical Society},
	doi = {10.1103/PhysRevB.58.R1703},
	url = {https://link.aps.org/doi/10.1103/PhysRevB.58.R1703}
}

@article{Kundu1999,
	title = {Exact Solution of Double $\ensuremath{\delta}$ Function Bose Gas through an Interacting Anyon Gas},
	author = {Kundu, Anjan},
	journal = {Phys. Rev. Lett.},
	volume = {83},
	issue = {7},
	pages = {1275--1278},
	numpages = {0},
	year = {1999},
	month = {Aug},
	publisher = {American Physical Society},
	doi = {10.1103/PhysRevLett.83.1275},
	url = {https://link.aps.org/doi/10.1103/PhysRevLett.83.1275}
}

@article{Zinner2015,
	title = {Strongly interacting mesoscopic systems of anyons in one dimension},
	author = {Zinner, N. T.},
	journal = {Phys. Rev. A},
	volume = {92},
	issue = {6},
	pages = {063634},
	numpages = {6},
	year = {2015},
	month = {Dec},
	publisher = {American Physical Society},
	doi = {10.1103/PhysRevA.92.063634},
	url = {https://link.aps.org/doi/10.1103/PhysRevA.92.063634}
}

@article{Batchelor2006,
	title = {One-Dimensional Interacting Anyon Gas: Low-Energy Properties and Haldane Exclusion Statistics},
	author = {Batchelor, M. T. and Guan, X.-W. and Oelkers, N.},
	journal = {Phys. Rev. Lett.},
	volume = {96},
	issue = {21},
	pages = {210402},
	numpages = {4},
	year = {2006},
	month = {Jun},
	publisher = {American Physical Society},
	doi = {10.1103/PhysRevLett.96.210402},
	url = {https://link.aps.org/doi/10.1103/PhysRevLett.96.210402}
}

@article{Girardeau2006,
	title = {Anyon-Fermion Mapping and Applications to Ultracold Gases in Tight Waveguides},
	author = {Girardeau, M. D.},
	journal = {Phys. Rev. Lett.},
	volume = {97},
	issue = {10},
	pages = {100402},
	numpages = {4},
	year = {2006},
	month = {Sep},
	publisher = {American Physical Society},
	doi = {10.1103/PhysRevLett.97.100402},
	url = {https://link.aps.org/doi/10.1103/PhysRevLett.97.100402}
}

@article{Hao2008,
	title = {Ground-state properties of one-dimensional anyon gases},
	author = {Hao, Yajiang and Zhang, Yunbo and Chen, Shu},
	journal = {Phys. Rev. A},
	volume = {78},
	issue = {2},
	pages = {023631},
	numpages = {6},
	year = {2008},
	month = {Aug},
	publisher = {American Physical Society},
	doi = {10.1103/PhysRevA.78.023631},
	url = {https://link.aps.org/doi/10.1103/PhysRevA.78.023631}
}

@article{Hao2009,
	title = {Ground-state properties of hard-core anyons in one-dimensional optical lattices},
	author = {Hao, Yajiang and Zhang, Yunbo and Chen, Shu},
	journal = {Phys. Rev. A},
	volume = {79},
	issue = {4},
	pages = {043633},
	numpages = {5},
	year = {2009},
	month = {Apr},
	publisher = {American Physical Society},
	doi = {10.1103/PhysRevA.79.043633},
	url = {https://link.aps.org/doi/10.1103/PhysRevA.79.043633}
}

@article{Hao2012,
	title = {Dynamical properties of hard-core anyons in one-dimensional optical lattices},
	author = {Hao, Yajiang and Chen, Shu},
	journal = {Phys. Rev. A},
	volume = {86},
	issue = {4},
	pages = {043631},
	numpages = {6},
	year = {2012},
	month = {Oct},
	publisher = {American Physical Society},
	doi = {10.1103/PhysRevA.86.043631},
	url = {https://link.aps.org/doi/10.1103/PhysRevA.86.043631}
}

@article{Greiter2009,
	title = {Statistical phases and momentum spacings for one-dimensional anyons},
	author = {Greiter, Martin},
	journal = {Phys. Rev. B},
	volume = {79},
	issue = {6},
	pages = {064409},
	numpages = {5},
	year = {2009},
	month = {Feb},
	publisher = {American Physical Society},
	doi = {10.1103/PhysRevB.79.064409},
	url = {https://link.aps.org/doi/10.1103/PhysRevB.79.064409}
}

@article{Arcila2016,
	title = {Critical points of the anyon-Hubbard model},
	author = {Arcila-Forero, J. and Franco, R. and Silva-Valencia, J.},
	journal = {Phys. Rev. A},
	volume = {94},
	issue = {1},
	pages = {013611},
	numpages = {9},
	year = {2016},
	month = {Jul},
	publisher = {American Physical Society},
	doi = {10.1103/PhysRevA.94.013611},
	url = {https://link.aps.org/doi/10.1103/PhysRevA.94.013611}
}

@article{Lange2017,
	title = {Anyonic Haldane Insulator in One Dimension},
	author = {Lange, Florian and Ejima, Satoshi and Fehske, Holger},
	journal = {Phys. Rev. Lett.},
	volume = {118},
	issue = {12},
	pages = {120401},
	numpages = {5},
	year = {2017},
	month = {Mar},
	publisher = {American Physical Society},
	doi = {10.1103/PhysRevLett.118.120401},
	url = {https://link.aps.org/doi/10.1103/PhysRevLett.118.120401}
}

@article{Zuo2018,
	title = {Statistically induced topological phase transitions in a one-dimensional superlattice anyon-Hubbard model},
	author = {Zuo, Zheng-Wei and Li, Guo-Ling and Li, Liben},
	journal = {Phys. Rev. B},
	volume = {97},
	issue = {11},
	pages = {115126},
	numpages = {7},
	year = {2018},
	month = {Mar},
	publisher = {American Physical Society},
	doi = {10.1103/PhysRevB.97.115126},
	url = {https://link.aps.org/doi/10.1103/PhysRevB.97.115126}
}

@article{Alexey2018,
	title = {Asymmetric Particle Transport and Light-Cone Dynamics Induced by Anyonic Statistics},
	author = {Liu, Fangli and Garrison, James R. and Deng, Dong-Ling and Gong, Zhe-Xuan and Gorshkov, Alexey V.},
	journal = {Phys. Rev. Lett.},
	volume = {121},
	issue = {25},
	pages = {250404},
	numpages = {7},
	year = {2018},
	month = {Dec},
	publisher = {American Physical Society},
	doi = {10.1103/PhysRevLett.121.250404},
	url = {https://link.aps.org/doi/10.1103/PhysRevLett.121.250404}
}

@article{Agarwala2019,
	title = {Statistics-tuned phases of pseudofermions in one dimension},
	author = {Agarwala, Adhip and Gupta, Gaurav Kumar and Shenoy, Vijay B. and Bhattacharjee, Subhro},
	journal = {Phys. Rev. B},
	volume = {99},
	issue = {16},
	pages = {165125},
	numpages = {15},
	year = {2019},
	month = {Apr},
	publisher = {American Physical Society},
	doi = {10.1103/PhysRevB.99.165125},
	url = {https://link.aps.org/doi/10.1103/PhysRevB.99.165125}
}

@article{Zhang2020,
	title = {Statistically related many-body localization in the one-dimensional anyon Hubbard model},
	author = {Zhang, Guo-Qing and Zhang, Dan-Wei and Li, Zhi and Wang, Z. D. and Zhu, Shi-Liang},
	journal = {Phys. Rev. B},
	volume = {102},
	issue = {5},
	pages = {054204},
	numpages = {9},
	year = {2020},
	month = {Aug},
	publisher = {American Physical Society},
	doi = {10.1103/PhysRevB.102.054204},
	url = {https://link.aps.org/doi/10.1103/PhysRevB.102.054204}
}

@article{Linhu2024,
	author = {Qin, Yi and Lee, Ching Hua and Li, Linhu},
	title = {Dynamical suppression of many-body non-Hermitian skin effect in anyonic systems},
	journal = {Commun. Phys.},
	volume = {8},
	pages = {18},
	ISSN = {2399-3650},
	DOI = {10.1038/s42005-025-01935-3},
	url = {https://doi.org/10.1038/s42005-025-01935-3},
	year = {2024},
	type = {Journal Article}
}

@article{Bonkhoff2025,
	title = {Anyonic Phase Transitions in the 1D Extended Hubbard Model with Fractional Statistics},
	author = {Bonkhoff, Martin and J\"agering, Kevin and Hu, Shijie and Pelster, Axel and Eggert, Sebastian and Schneider, Imke},
	journal = {Phys. Rev. Lett.},
	volume = {135},
	issue = {3},
	pages = {036601},
	numpages = {10},
	year = {2025},
	month = {Jul},
	publisher = {American Physical Society},
	doi = {10.1103/7n1c-vq2p},
	url = {https://link.aps.org/doi/10.1103/7n1c-vq2p}
}

@article{Keilmann2011,
	author = {Keilmann, Tassilo and Lanzmich, Simon and McCulloch, Ian and Roncaglia, Marco},
	title = {Statistically induced phase transitions and anyons in 1D optical lattices},
	journal = {Nat. Commun.},
	volume = {2},
	pages = {361},
	ISSN = {2041-1723},
	DOI = {10.1038/ncomms1353},
	url = {https://doi.org/10.1038/ncomms1353},
	year = {2011},
	type = {Journal Article}
}

@article{Greschner2015,
	title = {Anyon Hubbard Model in One-Dimensional Optical Lattices},
	author = {Greschner, Sebastian and Santos, Luis},
	journal = {Phys. Rev. Lett.},
	volume = {115},
	issue = {5},
	pages = {053002},
	numpages = {5},
	year = {2015},
	month = {Jul},
	publisher = {American Physical Society},
	doi = {10.1103/PhysRevLett.115.053002},
	url = {https://link.aps.org/doi/10.1103/PhysRevLett.115.053002}
}

@article{Christoph2016,
  	title = {Floquet Realization and Signatures of One-Dimensional Anyons in an Optical Lattice},
  	author = {Str\"ater, Christoph and Srivastava, Shashi C. L. and Eckardt, Andr\'e},
  	journal = {Phys. Rev. Lett.},
  	volume = {117},
  	issue = {20},
  	pages = {205303},
  	numpages = {6},
  	year = {2016},
  	month = {Nov},
  	publisher = {American Physical Society},
  	doi = {10.1103/PhysRevLett.117.205303},
  	url = {https://link.aps.org/doi/10.1103/PhysRevLett.117.205303}
  }

@article{Yuan2017,
	title = {Creating anyons from photons using a nonlinear resonator lattice subject to dynamic modulation},
	author = {Yuan, Luqi and Xiao, Meng and Xu, Shanshan and Fan, Shanhui},
	journal = {Phys. Rev. A},
	volume = {96},
	issue = {4},
	pages = {043864},
	numpages = {6},
	year = {2017},
	month = {Oct},
	publisher = {American Physical Society},
	doi = {10.1103/PhysRevA.96.043864},
	url = {https://link.aps.org/doi/10.1103/PhysRevA.96.043864}
}

@article{Frederik2019,
	author = {Görg, Frederik and Sandholzer, Kilian and Minguzzi, Joaquín and Desbuquois, Rémi and Messer, Michael and
	Esslinger, Tilman},
	title = {Realization of density-dependent Peierls phases to engineer quantized gauge fields coupled to ultracold matter},
	journal = {Nat. Phys.},
	volume = {15},
	pages = {1161-1167},
	ISSN = {1745-2481},
	DOI = {10.1038/s41567-019-0615-4},
	url = {https://doi.org/10.1038/s41567-019-0615-4},
	year = {2019},
	type = {Journal Article}
}

@article{Lienhard2020,
	title = {Realization of a Density-Dependent Peierls Phase in a Synthetic, Spin-Orbit Coupled Rydberg System},
	author = {Lienhard, Vincent and Scholl, Pascal and Weber, Sebastian and Barredo, Daniel and de L\'es\'eleuc, Sylvain and Bai, Rukmani and Lang, Nicolai and Fleischhauer, Michael and B\"uchler, Hans Peter and Lahaye, Thierry and Browaeys, Antoine},
	journal = {Phys. Rev. X},
	volume = {10},
	issue = {2},
	pages = {021031},
	numpages = {11},
	year = {2020},
	month = {May},
	publisher = {American Physical Society},
	doi = {10.1103/PhysRevX.10.021031},
	url = {https://link.aps.org/doi/10.1103/PhysRevX.10.021031}
}

@Article{Greiner2024,
  author  = {Joyce Kwan and Perrin Segura and Yanfei Li and Sooshin Kim and Alexey V. Gorshkov and André Eckardt and Brice Bakkali-Hassani and Markus Greiner},
  journal = {Science},
  title   = {Realization of one-dimensional anyons with arbitrary statistical phase},
  year    = {2024},
  number  = {6725},
  pages   = {1055-1060},
  volume  = {386},
  doi     = {10.1126/science.adi3252},
  url     = {https://doi/abs/10.1126/science.adi3252},
}

@Article{Dhar2025,
  author  = {Dhar, Sudipta and Wang, Botao and Horvath, Milena and Vashisht, Amit and Zeng, Yi and Zvonarev, Mikhail B. and Goldman, Nathan and Guo, Yanliang and Landini, Manuele and N\"agerl, Hanns-Christoph},
  journal = {Nature},
  title   = {Observing anyonization of bosons in a quantum gas},
  year    = {2025},
  issn    = {1476-4687},
  number  = {8066},
  pages   = {53-57},
  volume  = {642},
  doi     = {10.1038/s41586-025-09016-9},
  type    = {Journal Article},
  url     = {https://doi.org/10.1038/s41586-025-09016-9},
}

@article{Dalibard2011,
	title = {Colloquium: Artificial gauge potentials for neutral atoms},
	author = {Dalibard, Jean and Gerbier, Fabrice and Juzeli\ifmmode \bar{u}\else \={u}\fi{}nas, Gediminas and \"Ohberg, Patrik},
	journal = {Rev. Mod. Phys.},
	volume = {83},
	issue = {4},
	pages = {1523--1543},
	numpages = {0},
	year = {2011},
	month = {Nov},
	publisher = {American Physical Society},
	doi = {10.1103/RevModPhys.83.1523},
	url = {https://link.aps.org/doi/10.1103/RevModPhys.83.1523}
}

@article{Goldman2014,
	doi = {10.1088/0034-4885/77/12/126401},
	url = {https://dx.doi.org/10.1088/0034-4885/77/12/126401},
	year = {2014},
	month = {nov},
	publisher = {IOP Publishing},
	volume = {77},
	number = {12},
	pages = {126401},
	author = {N Goldman and G Juzeliūnas and P Öhberg and I B Spielman},
	title = {Light-induced gauge fields for ultracold atoms},
	journal = {Reports on Progress in Physics}
}

@article{Lin2009,
	author={Lin, Y-J and Compton, Rob L and Jim{\'e}nez-Garc{\'\i}a, Karina and Porto, James V and Spielman, Ian B},
	title={Synthetic magnetic fields for ultracold neutral atoms},
	journal={Nature},
	volume={462},
	number={7273},
	pages={628-632},
	ISSN = {1476-4687},
	DOI = {10.1038/nature08609},
	year={2009},
	publisher={Nature Publishing Group UK London}
}

@article{Aidelsburger2013,
	title = {Realization of the Hofstadter Hamiltonian with Ultracold Atoms in Optical Lattices},
	author = {Aidelsburger, M. and Atala, M. and Lohse, M. and Barreiro, J. T. and Paredes, B. and Bloch, I.},
	journal = {Phys. Rev. Lett.},
	volume = {111},
	issue = {18},
	pages = {185301},
	numpages = {5},
	year = {2013},
	month = {Oct},
	publisher = {American Physical Society},
	doi = {10.1103/PhysRevLett.111.185301},
	url = {https://link.aps.org/doi/10.1103/PhysRevLett.111.185301}
}

@article{Miyake2013,
	title = {Realizing the Harper Hamiltonian with Laser-Assisted Tunneling in Optical Lattices},
	author = {Miyake, Hirokazu and Siviloglou, Georgios A. and Kennedy, Colin J. and Burton, William Cody and Ketterle, Wolfgang},
	journal = {Phys. Rev. Lett.},
	volume = {111},
	issue = {18},
	pages = {185302},
	numpages = {5},
	year = {2013},
	month = {Oct},
	publisher = {American Physical Society},
	doi = {10.1103/PhysRevLett.111.185302},
	url = {https://link.aps.org/doi/10.1103/PhysRevLett.111.185302}
}

@article{Yan2022,
	title = {Aharonov-Bohm Caging and Inverse Anderson Transition in Ultracold Atoms},
	author = {Li, Hang and Dong, Zhaoli and Longhi, Stefano and Liang, Qian and Xie, Dizhou and Yan, Bo},
	journal = {Phys. Rev. Lett.},
	volume = {129},
	issue = {22},
	pages = {220403},
	numpages = {6},
	year = {2022},
	month = {Nov},
	publisher = {American Physical Society},
	doi = {10.1103/PhysRevLett.129.220403},
	url = {https://link.aps.org/doi/10.1103/PhysRevLett.129.220403}
}

@article{Yan2025EngineeringTC,
    author = {Li, Hang and Liang, Qian and Dong, Zhaoli and Wang, Hongru and Yi, Wei and Pan, Jian-Song and Yan, Bo},
    title = {Engineering topological chiral transport in a flat-band lattice of ultracold atoms},
    journal = {Light: Science \& Applications},
    volume = {14},
    pages = {326},
    ISSN = {2047-7538},
    DOI = {10.1038/s41377-025-02025-3},
    url = {https://doi.org/10.1038/s41377-025-02025-3},
    year = {2025},
    type = {Journal Article}
}

@article{Goldman2016,
	author = {Goldman, N. and
	Budich, J. C. and
	Zoller, P.},
	title = {Topological quantum matter with ultracold gases in optical lattices},
	journal = {Nat. Phys.},
	volume = {12},
	number = {7},
	pages = {639-645},
	ISSN = {1745-2481},
	DOI = {10.1038/nphys3803},
	url = {https://doi.org/10.1038/nphys3803},
	year = {2016},
}

@article{Cooper2019,
	title = {Topological bands for ultracold atoms},
	author = {Cooper, N. R. and Dalibard, J. and Spielman, I. B.},
	journal = {Rev. Mod. Phys.},
	volume = {91},
	issue = {1},
	pages = {015005},
	numpages = {55},
	year = {2019},
	month = {Mar},
	publisher = {American Physical Society},
	doi = {10.1103/RevModPhys.91.015005},
	url = {https://link.aps.org/doi/10.1103/RevModPhys.91.015005}
}

@article{Dariol2014,
	title = {Chiral ladders and the edges of quantum Hall insulators},
	author = {H\"ugel, Dario and Paredes, Bel\'en},
	journal = {Phys. Rev. A},
	volume = {89},
	issue = {2},
	pages = {023619},
	numpages = {7},
	year = {2014},
	month = {Feb},
	publisher = {American Physical Society},
	doi = {10.1103/PhysRevA.89.023619},
	url = {https://link.aps.org/doi/10.1103/PhysRevA.89.023619}
}

@article{Piraud2015,
	title = {Vortex and Meissner phases of strongly interacting bosons on a two-leg ladder},
	author = {Piraud, M. and Heidrich-Meisner, F. and McCulloch, I. P. and Greschner, S. and Vekua, T. and Schollw\"ock, U.},
	journal = {Phys. Rev. B},
	volume = {91},
	issue = {14},
	pages = {140406},
	numpages = {5},
	year = {2015},
	month = {Apr},
	publisher = {American Physical Society},
	doi = {10.1103/PhysRevB.91.140406},
	url = {https://link.aps.org/doi/10.1103/PhysRevB.91.140406}
}

@article{Zheng2017,
	title = {Chiral Bloch oscillation and nontrivial topology in a ladder lattice with magnetic flux},
	author = {Zheng, Yi and Feng, Shiping and Yang, Shi-Jie},
	journal = {Phys. Rev. A},
	volume = {96},
	issue = {6},
	pages = {063613},
	numpages = {8},
	year = {2017},
	month = {Dec},
	publisher = {American Physical Society},
	doi = {10.1103/PhysRevA.96.063613},
	url = {https://link.aps.org/doi/10.1103/PhysRevA.96.063613}
}

@article{Bryce2017,
	title={Diffusive and arrested transport of atoms under tailored disorder},
	author={An, Fangzhao Alex and
	Meier, Eric J. and
	Gadway, Bryce},
	journal = {Nat. Commun.},
	volume = {8},
	number = {1},
	pages = {325},
	ISSN = {2041-1723},
	DOI = {10.1038/s41467-017-00387-w},
	url = {https://doi.org/10.1038/s41467-017-00387-w},
	year = {2017},
}

@article{Li2020,
	author = {Li, Y. and Cai, H. and Wang, D. W. and Li, L. and Yuan, J. and Li, W.},
	title = {Many-Body Chiral Edge Currents and Sliding Phases of Atomic Spin Waves in Momentum-Space Lattice},
	journal = {Phys Rev Lett},
	volume = {124},
	number = {14},
	pages = {140401},
	ISSN = {1079-7114 (Electronic)
	0031-9007 (Linking)},
	DOI = {10.1103/PhysRevLett.124.140401},
	url = {https://www.ncbi.nlm.nih.gov/pubmed/32338979},
	year = {2020},
	type = {Journal Article}
}

@article{Bryce2017Direct,
  	author  = {Fangzhao Alex An and Eric J. Meier and Bryce Gadway},
  	journal = {Science Advances},
  	title   = {Direct observation of chiral currents and magnetic reflection in atomic flux lattices},
  	year    = {2017},
  	number  = {4},
  	pages   = {e1602685},
  	volume  = {3},
  	doi     = {10.1126/sciadv.1602685},
}

@article{dutt2020single,
	title={A single photonic cavity with two independent physical synthetic dimensions},
	author={Dutt, Avik and Lin, Qian and Yuan, Luqi and Minkov, Momchil and Xiao, Meng and Fan, Shanhui},
	journal={Science},
	volume={367},
	number={6473},
	pages={59--64},
	year={2020},
	publisher={American Association for the Advancement of Science},
	doi = {10.1126/science.aaz3071},
	URL = {https://www.science.org/doi/abs/10.1126/science.aaz3071}
}

@article{Suotang2022,
	title={Atom-optically synthetic gauge fields for a noninteracting Bose gas},
	author={Li, Yuqing and
	Zhang, Jiahui and
	Wang, Yunfei and
	Du, Huiying and
	Wu, Jizhou and
	Liu, Wenliang and
	Mei, Feng and
	Ma, Jie and
	Xiao, Liantuan and
	Jia, Suotang},
	journal = {Light: Science \& Applications},
	volume = {11},
	number = {1},
	pages = {13},
	ISSN = {2047-7538},
	DOI = {10.1038/s41377-021-00702-7},
	url = {https://doi.org/10.1038/s41377-021-00702-7},
	year = {2022},
}

@article{Wu2022,
	title = {Flux-controlled skin effect and topological transition in a dissipative two-leg ladder model},
	author = {Wu, Chaohua and Yang, Zhesen and Tang, Jiangshan and Liu, Ni and Chen, Gang},
	journal = {Phys. Rev. A},
	volume = {106},
	issue = {6},
	pages = {062206},
	numpages = {9},
	year = {2022},
	month = {Dec},
	publisher = {American Physical Society},
	doi = {10.1103/PhysRevA.106.062206},
	url = {https://link.aps.org/doi/10.1103/PhysRevA.106.062206}
}

@article{Chen2024,
	title = {Antichiral and trap-skin dynamics in a nonreciprocal bosonic two-leg ladder with artificial magnetic flux},
	author = {Chen, Rui-Jie and Zhang, Guo-Qing and Li, Zhi and Zhang, Dan-Wei},
	journal = {Phys. Rev. A},
	volume = {110},
	issue = {4},
	pages = {043311},
	numpages = {8},
	year = {2024},
	month = {Oct},
	publisher = {American Physical Society},
	doi = {10.1103/PhysRevA.110.043311},
	url = {https://link.aps.org/doi/10.1103/PhysRevA.110.043311}
}

@article{Ye2025,
	title={Observing non-Hermiticity induced chirality breaking in a synthetic Hall ladder},
	author={Ye, Rui and He, Yanyan and Li, Guangzhen and Wang, Luojia and Wu, Xiaoxiong and Qiao, Xin and Zheng, Yuanlin and Jin, Liang and Wang, Da-Wei and Yuan, Luqi and Chen Xianfeng},
	journal={Light: Science \& Applications},
	volume={14},
	number={1},
	pages={39},
	DOI = {10.1038/s41377-024-01700-1},
	url = {https://doi.org/10.1038/s41377-024-01700-1},
	year={2025},
	publisher={Nature Publishing Group UK London}	
}

@article{Ronzheimer2013,
	title = {Expansion Dynamics of Interacting Bosons in Homogeneous Lattices in One and Two Dimensions},
	author = {Ronzheimer, J. P. and Schreiber, M. and Braun, S. and Hodgman, S. S. and Langer, S. and McCulloch, I. P. and Heidrich-Meisner, F. and Bloch, I. and Schneider, U.},
	journal = {Phys. Rev. Lett.},
	volume = {110},
	issue = {20},
	pages = {205301},
	numpages = {6},
	year = {2013},
	month = {May},
	publisher = {American Physical Society},
	doi = {10.1103/PhysRevLett.110.205301},
	url = {https://link.aps.org/doi/10.1103/PhysRevLett.110.205301}
}

@Article{Dai2017,
  author    = {Dai, Han-Ning and Yang, Bing and Reingruber, Andreas and Sun, Hui and Xu, Xiao-Fan and Chen, Yu-Ao and Yuan, Zhen-Sheng and Pan, Jian-Wei},
  journal   = {Nature Physics},
  title     = {Four-body ring-exchange interactions and anyonic statistics within a minimal toric-code Hamiltonian},
  year      = {2017},
  issn      = {1745-2481},
  month     = aug,
  number    = {12},
  pages     = {1195--1200},
  volume    = {13},
  doi       = {10.1038/nphys4243},
  publisher = {Springer Science and Business Media LLC},
}

@Article{Paredes2008,
  author    = {Paredes, Belén and Bloch, Immanuel},
  journal   = {Physical Review A},
  title     = {Minimum instances of topological matter in an optical plaquette},
  year      = {2008},
  issn      = {1094-1622},
  month     = feb,
  number    = {2},
  pages     = {023603},
  volume    = {77},
  doi       = {10.1103/physreva.77.023603},
  publisher = {American Physical Society (APS)},
}

@Article{Greschner2018,
  author    = {Greschner, Sebastian and Cardarelli, Lorenzo and Santos, Luis},
  journal   = {Physical Review A},
  title     = {Probing the exchange statistics of one-dimensional anyon models},
  year      = {2018},
  issn      = {2469-9934},
  month     = may,
  number    = {5},
  pages     = {053605},
  volume    = {97},
  doi       = {10.1103/physreva.97.053605},
  publisher = {American Physical Society (APS)},
}

@Article{Cardarelli2016,
  author    = {Cardarelli, Lorenzo and Greschner, Sebastian and Santos, Luis},
  journal   = {Physical Review A},
  title     = {Engineering interactions and anyon statistics by multicolor lattice-depth modulations},
  year      = {2016},
  issn      = {2469-9934},
  month     = aug,
  number    = {2},
  pages     = {023615},
  volume    = {94},
  doi       = {10.1103/physreva.94.023615},
  publisher = {American Physical Society (APS)},
}

@Article{Greschner2014,
  author    = {Greschner, S. and Sun, G. and Poletti, D. and Santos, L.},
  journal   = {Physical Review Letters},
  title     = {Density-Dependent Synthetic Gauge Fields Using Periodically Modulated Interactions},
  year      = {2014},
  issn      = {1079-7114},
  month     = nov,
  number    = {21},
  pages     = {215303},
  volume    = {113},
  doi       = {10.1103/physrevlett.113.215303},
  publisher = {American Physical Society (APS)},
}

@Article{Yu2017,
  author    = {Yu, Jinlong and Sun, Ning and Zhai, Hui},
  journal   = {Physical Review Letters},
  title     = {Symmetry Protected Dynamical Symmetry in the Generalized Hubbard Models},
  year      = {2017},
  issn      = {1079-7114},
  month     = nov,
  number    = {22},
  pages     = {225302},
  volume    = {119},
  doi       = {10.1103/physrevlett.119.225302},
  publisher = {American Physical Society (APS)},
}

@Article{Schneider2012,
  author    = {Schneider, Ulrich and Hackermüller, Lucia and Ronzheimer, Jens Philipp and Will, Sebastian and Braun, Simon and Best, Thorsten and Bloch, Immanuel and Demler, Eugene and Mandt, Stephan and Rasch, David and Rosch, Achim},
  journal   = {Nature Physics},
  title     = {Fermionic transport and out-of-equilibrium dynamics in a homogeneous Hubbard model with ultracold atoms},
  year      = {2012},
  issn      = {1745-2481},
  month     = jan,
  number    = {3},
  pages     = {213--218},
  volume    = {8},
  doi       = {10.1038/nphys2205},
  publisher = {Springer Science and Business Media LLC},
}

@Article{BWang2025,
  author    = {Wang, Botao and Vashisht, Amit and Guo, Yanliang and Dhar, Sudipta and Landini, Manuele and N\"agerl, Hanns-Christoph and Goldman, Nathan},
  journal   = {Phys. Rev. Lett.},
  title     = {Anyonization of Bosons in One Dimension: An Effective Swap Model},
  year      = {2025},
  month     = {Dec},
  pages     = {253403},
  volume    = {135},
  doi       = {10.1103/2np8-mp39},
  issue     = {25},
  numpages  = {8},
  publisher = {American Physical Society},
  url       = {https://link.aps.org/doi/10.1103/2np8-mp39},
}

@Article{JWang2025a,
  author    = {Wang, Junsen and Sun, Xiangxiang and Zheng, Wei},
  journal   = {Phys. Rev. Res.},
  title     = {Quantum simulation with gauge fixing: From Ising lattice gauge theory to dynamical flux model},
  year      = {2025},
  month     = {Mar},
  pages     = {013311},
  volume    = {7},
  doi       = {10.1103/PhysRevResearch.7.013311},
  issue     = {1},
  numpages  = {13},
  publisher = {American Physical Society},
  url       = {https://link.aps.org/doi/10.1103/PhysRevResearch.7.013311},
}

@Article{Santachiara2008,
  author  = {Santachiara, Raoul and Calabrese, Pasquale},
  journal = {Journal of Statistical Mechanics: Theory and Experiment},
  title   = {One-particle density matrix and momentum distribution function of one-dimensional anyon gases},
  year    = {2008},
  month   = {jun},
  number  = {06},
  pages   = {P06005},
  volume  = {2008},
  doi     = {10.1088/1742-5468/2008/06/P06005},
  url     = {https://doi.org/10.1088/1742-5468/2008/06/P06005},
}

@Article{Scopa2020,
  author    = {Scopa, Stefano and Piroli, Lorenzo and Calabrese, Pasquale},
  journal   = {Journal of Statistical Mechanics: Theory and Experiment},
  title     = {One-particle density matrix of a trapped Lieb–Liniger anyonic gas},
  year      = {2020},
  month     = {sep},
  number    = {9},
  pages     = {093103},
  volume    = {2020},
  doi       = {10.1088/1742-5468/abaed1},
  publisher = {IOP Publishing and SISSA},
  url       = {https://doi.org/10.1088/1742-5468/abaed1},
}

@Article{OIP2020,
  author    = {P\^a\ifmmode \mbox{\c{t}}\else \c{t}\fi{}u, Ovidiu I.},
  journal   = {Phys. Rev. A},
  title     = {Nonequilibrium dynamics of the anyonic Tonks-Girardeau gas at finite temperature},
  year      = {2020},
  month     = {Oct},
  pages     = {043303},
  volume    = {102},
  doi       = {10.1103/PhysRevA.102.043303},
  issue     = {4},
  numpages  = {21},
  publisher = {American Physical Society},
  url       = {https://link.aps.org/doi/10.1103/PhysRevA.102.043303},
}

@Article{OIP2022,
  author    = {P\^a\ifmmode \mbox{\c{t}}\else \c{t}\fi{}u, Ovidiu I.},
  journal   = {Phys. Rev. A},
  title     = {Exact spectral function of the Tonks-Girardeau gas at finite temperature},
  year      = {2022},
  month     = {Nov},
  pages     = {053306},
  volume    = {106},
  doi       = {10.1103/PhysRevA.106.053306},
  issue     = {5},
  numpages  = {19},
  publisher = {American Physical Society},
  url       = {https://link.aps.org/doi/10.1103/PhysRevA.106.053306},
}

@Article{OIP2025,
  author    = {P\^a\ifmmode \mbox{\c{t}}\else \c{t}\fi{}u, Ovidiu I.},
  journal   = {Phys. Rev. A},
  title     = {Universal properties and dynamical bosonization of strongly interacting one-dimensional anyons},
  year      = {2025},
  month     = {Dec},
  pages     = {063312},
  volume    = {112},
  doi       = {10.1103/h3hs-g475},
  issue     = {6},
  numpages  = {16},
  publisher = {American Physical Society},
  url       = {https://link.aps.org/doi/10.1103/h3hs-g475},
}

@Article{Calabrese2009,
  author    = {Calabrese, Pasquale and Santachiara, Raoul},
  journal   = {Journal of Statistical Mechanics: Theory and Experiment},
  title     = {Off-diagonal correlations in one-dimensional anyonic models: a replica approach},
  year      = {2009},
  issn      = {1742-5468},
  month     = mar,
  number    = {03},
  pages     = {P03002},
  volume    = {2009},
  doi       = {10.1088/1742-5468/2009/03/p03002},
  publisher = {IOP Publishing},
}

@Article{Fishman2022,
  author    = {Fishman, Matthew and White, Steven and Stoudenmire, Edwin},
  journal   = {SciPost Phys. Codebases},
  title     = {The ITensor Software Library for Tensor Network Calculations},
  year      = {2022},
  month     = aug,
  doi       = {10.21468/scipostphyscodeb.4},
  publisher = {Stichting SciPost},
}

@Article{Zhu2011,
  author    = {Zhu, Shi-Liang and Shao, L.-B. and Wang, Z. D. and Duan, L.-M.},
  journal   = {Phys. Rev. Lett.},
  title     = {Probing Non-Abelian Statistics of Majorana Fermions in Ultracold Atomic Superfluid},
  year      = {2011},
  month     = {Mar},
  pages     = {100404},
  volume    = {106},
  doi       = {10.1103/PhysRevLett.106.100404},
  issue     = {10},
  numpages  = {4},
  publisher = {American Physical Society},
  url       = {https://link.aps.org/doi/10.1103/PhysRevLett.106.100404},
}

@Article{Zhu2006,
  author    = {Zhu, Shi-Liang and Fu, Hao and Wu, C.-J. and Zhang, S.-C. and Duan, L.-M.},
  journal   = {Phys. Rev. Lett.},
  title     = {Spin Hall Effects for Cold Atoms in a Light-Induced Gauge Potential},
  year      = {2006},
  month     = {Dec},
  pages     = {240401},
  volume    = {97},
  doi       = {10.1103/PhysRevLett.97.240401},
  issue     = {24},
  numpages  = {4},
  publisher = {American Physical Society},
  url       = {https://link.aps.org/doi/10.1103/PhysRevLett.97.240401},
}

@Article{GQZhang2021,
  author    = {Zhang, Guo-Qing and Tang, Ling-Zhi and Zhang, Ling-Feng and Zhang, Dan-Wei and Zhu, Shi-Liang},
  journal   = {Phys. Rev. B},
  title     = {Connecting topological Anderson and Mott insulators in disordered interacting fermionic systems},
  year      = {2021},
  month     = {Oct},
  pages     = {L161118},
  volume    = {104},
  doi       = {10.1103/PhysRevB.104.L161118},
  issue     = {16},
  numpages  = {6},
  publisher = {American Physical Society},
  url       = {https://link.aps.org/doi/10.1103/PhysRevB.104.L161118},
}

@Article{YLChen2020,
  author    = {Chen, Yu-Lian and Zhang, Guo-Qing and Zhang, Dan-Wei and Zhu, Shi-Liang},
  journal   = {Phys. Rev. A},
  title     = {Simulating bosonic Chern insulators in one-dimensional optical superlattices},
  year      = {2020},
  month     = {Jan},
  pages     = {013627},
  volume    = {101},
  doi       = {10.1103/PhysRevA.101.013627},
  issue     = {1},
  numpages  = {9},
  publisher = {American Physical Society},
  url       = {https://link.aps.org/doi/10.1103/PhysRevA.101.013627},
}

@article{Nayak2008,
	title = {Non-Abelian anyons and topological quantum computation},
	author = {Nayak, Chetan and Simon, Steven H. and Stern, Ady and Freedman, Michael and Das Sarma, Sankar},
	journal = {Rev. Mod. Phys.},
	volume = {80},
	issue = {3},
	pages = {1083--1159},
	numpages = {0},
	year = {2008},
	month = {Sep},
	publisher = {American Physical Society},
	doi = {10.1103/RevModPhys.80.1083},
	url = {https://link.aps.org/doi/10.1103/RevModPhys.80.1083}
}

@article{DW2018,
	author = {Dan-Wei Zhang and Yan-Qing Zhu and Y. X. Zhao and Hui Yan and Shi-Liang Zhu},
	title = {Topological quantum matter with cold atoms},
	journal = {Advances in Physics},
	volume = {67},
	number = {4},
	pages = {253--402},
	year = {2018},
	publisher = {Taylor \& Francis},
	doi = {10.1080/00018732.2019.1594094},	
	url = {https://doi.org/10.1080/00018732.2019.1594094},
}

@article{Atala2014,
	author = {Atala, Marcos and Aidelsburger, Monika and Lohse, Michael and Barreiro, Julio T. and Paredes, Belén and Bloch, Immanuel},
	title = {Observation of chiral currents with ultracold atoms in bosonic ladders},
	journal = {Nat. Phys.},
	volume = {10},
	number = {8},
	pages = {588-593},
	ISSN = {1745-2473
	1745-2481},
	DOI = {10.1038/nphys2998},
	year = {2014},
	type = {Journal Article}
}

@article{Tai2017,
	author = {Tai, M. E. and Lukin, A. and Rispoli, M. and Schittko, R. and Menke, T. and Dan, Borgnia and Preiss, P. M. and Grusdt, F. and Kaufman, A. M. and Greiner, M.},
	title = {Microscopy of the interacting Harper-Hofstadter model in the two-body limit},
	journal = {Nature},
	volume = {546},
	number = {7659},
	pages = {519-523},
	ISSN = {1476-4687 (Electronic)
	0028-0836 (Linking)},
	DOI = {10.1038/nature22811},
	url = {https://doi.org/10.1038/nature22811},
	year = {2017},
	type = {Journal Article}
}

@article{Yan2024,
	author = {Liang, Qian and Dong, Zhaoli and Pan, Jian-Song and Wang, Hongru and Li, Hang and Yang, Zhaoju and Yi, Wei and Yan, Bo},
	title = {Chiral dynamics of ultracold atoms under a tunable SU(2) synthetic gauge field},
	journal = {Nat. Phys.},
	volume = {20},
	pages = {1738-1743},
	ISSN = {1745-2481},
	DOI = {10.1038/s41567-024-02644-4},
	url = {https://doi.org/10.1038/s41567-024-02644-4},
	year = {2024},
	type = {Journal Article}
}

@article{Giri2023,
	title = {Flux-induced reentrant dynamics in the quantum walk of interacting bosons},
	author = {Giri, Mrinal Kanti and Paul, Biswajit and Mishra, Tapan},
	journal = {Phys. Rev. A},
	volume = {108},
	issue = {6},
	pages = {063319},
	numpages = {6},
	year = {2023},
	month = {Dec},
	publisher = {American Physical Society},
	doi = {10.1103/PhysRevA.108.063319},
	url = {https://link.aps.org/doi/10.1103/PhysRevA.108.063319}
}

@Article{DWZhang2020,
  author    = {Zhang, Dan-Wei and Chen, Yu-Lian and Zhang, Guo-Qing and Lang, Li-Jun and Li, Zhi and Zhu, Shi-Liang},
  journal   = {Phys. Rev. B},
  title     = {Skin superfluid, topological Mott insulators, and asymmetric dynamics in an interacting non-Hermitian Aubry-Andr\'e-Harper model},
  year      = {2020},
  month     = {Jun},
  pages     = {235150},
  volume    = {101},
  doi       = {10.1103/PhysRevB.101.235150},
  issue     = {23},
  numpages  = {9},
  publisher = {American Physical Society},
  url       = {https://link.aps.org/doi/10.1103/PhysRevB.101.235150},
}
\end{document}